\documentclass[reqno]{amsart}

\usepackage{booktabs} 
\usepackage{IEEEtrantools}
\usepackage{amssymb,latexsym,amsfonts,amsmath}
\usepackage{graphicx}
\usepackage{mathrsfs}
\usepackage{dsfont}

\usepackage{tikz}
\usetikzlibrary{calc,shapes,arrows}

\usepackage{algorithm}
\usepackage{algorithmic}

\usepackage{xspace}

\newtheorem{theorem}{Theorem}[section]
\newtheorem{lemma}[theorem]{Lemma}

\newtheorem{proposition}[theorem]{Proposition}

\newtheorem{definition}[theorem]{Definition}

\newtheorem{remark}[theorem]{Remark}
\newtheorem{assumption}{Assumption}
\numberwithin{equation}{section}

\usepackage{multicol}
\usepackage{enumerate}
\usepackage{latexsym}
\usepackage{mathrsfs}
\usepackage{dsfont}
\usepackage{textcomp}
\usepackage{xcolor}

\begin{document}
	
\begin{abstract}
In this paper, we propose a compositional framework for the synthesis of safety controllers for networks of partially-observed discrete-time stochastic control systems (a.k.a. continuous-space POMDPs). Given an estimator, we utilize a discretization-free approach to synthesize controllers ensuring safety specifications over finite-time horizons. The proposed framework is based on a notion of so-called \textit{local control barrier functions} computed for subsystems in two different ways.  In the first scheme, no prior knowledge of estimation accuracy is needed. The second framework utilizes a probability bound on the estimation accuracy using a notion of so called \textit{stochastic simulation functions}. In both proposed schemes, we drive sufficient small-gain type conditions in order to compositionally construct control barrier functions for interconnected POMDPs using local barrier functions computed for subsystems. Leveraging compositionality results, the constructed control barrier functions enable us to compute  lower bounds on the probabilities that the interconnected POMDPs avoid certain unsafe regions in finite-time horizons. We demonstrate the effectiveness
of our proposed approaches by applying them to an adaptive cruise control problem.
\end{abstract}

\title[Compositional Construction of Safety Controllers for Networks of Continuous-Space POMDPs]{Compositional Construction of Safety Controllers for Networks of Continuous-Space POMDPs}

\author{Niloofar Jahanshahi$^{1}$}
\author{Abolfazl Lavaei$^{2}$}
\author{Majid Zamani$^{3,1}$}
\address{$^1$Department of Computer Science, LMU Munich, Germany}
\email{niloofar.jahanshahi@lmu.de}
\address{$^2$Institute for Dynamic Systems and Control, ETH Zurich, Switzerland}
\email{alavaei@ethz.ch}
\address{$^3$Department of Computer Science, University of Colorado Boulder, USA}
\email{majid.zamani@colorado.edu}
\maketitle

	\section{Introduction}
	Large-scale stochastic systems have received significant attentions in the past few years due to their broad applications in modeling many engineering systems such as power grids, road traffic networks and industrial control systems to name a few. Guaranteeing safety and reliability of such complex systems
	in a formal as well as time- and cost-effective way has always been very challenging. In the past few years, formal verification and synthesis of controllers against safety specifications have gained considerable attentions among both control engineers and computer scientists. In this respect, abstraction-based techniques have been widely employed for the formal synthesis of safety controllers~\cite{lygeros1996hierarchical, lavaei2020compositional, tran2017order}. However, those approaches rely on the state and input set discretization and consequently suffer severely from the \emph{curse of dimensionality}: computational complexity exponentially grows with the dimension of the system. In order to overcome this difficulty, compositional techniques have been introduced in the past few years to construct  finite abstractions of interconnected systems based on abstractions of smaller subsystems~\cite{SAM17,lavaei2018CDCJ,lavaei2019HSCC_J,AmyJournal2020,lavaei2020compositional,AmyIFAC2020,lavaei2019NAHS,lavaei2019LSS,lavaei2017HSCC,lavaei2019NAHS1,lavaei2019Thesis,Lavaei_Survey}.
	
	As another promising alternative, \emph{discretization-free} approaches based on \emph{control barrier functions} have been introduced in the past decade~\cite{prajna2007framework,ames2016control, nejati2020compositional, ames2019control, jagtap2020compositional,AmyIFAC12020,nejati2020compositional1,mahathi_IFAC2020,anand2021smallgain}. Unfortunately, all above-mentioned literatures on both discretization and discretization-free techniques assume that full state information is available for the sake of controller synthesis which is not the case in many practical applications. Taking this limitation into account, the work in \cite{clark2019control} studies a controller synthesis scheme for stochastic systems with incomplete information by assuming a priori knowledge of control barrier functions. Given an estimator with a probabilistic guarantee on the accuracy of estimations, \cite{CBF1NPM} studies the controller synthesis problem for partially-observed stochastic systems and proposes a lower bound for the probability of satisfaction of safety specifications over finite-time horizons. A synthesis framework based on control barrier functions for partially-observed jump diffusion systems enforcing complex properties expressed by deterministic finite automata is recently proposed in~\cite{jahanshahi2020synthesis} in which a prior knowledge of the estimation accuracy is not required anymore. 
	
	The proposed techniques in the above-mentioned literature on partially observed systems assume that control barrier functions have a certain parametric form, such as polynomial, and search for their corresponding coefficients under certain assumptions. Although it may be easy to search for those functions for lower-dimensional systems via existing tools, it is computationally very expensive (if not impossible) to compute them for large-scale interconnected systems. Motivated by this challenge, we propose here a compositional approach for the construction of control barrier functions for partially-observed discrete-time stochastic control systems (a.k.a. POMDPs). To the best of our knowledge, this paper is the first to develop a compositional controller synthesis scheme for networks of  POMDPs based on barrier functions. By driving small-gain type conditions, we compositionally construct a control barrier function for the interconnected POMDP based on local  barrier functions of subsystems. Accordingly, by leveraging the constructed barrier function and the corresponding controller, we compute a lower bound on the probability that the interconnected POMDP avoids an unsafe region over a finite-time horizon.
	
	Particularly, we propose two distinct approaches for the construction of control barrier functions. In the first one, local control barrier functions are defined over augmented systems consisting of subsystems and their estimators. This formulation makes it possible to search for local control barrier functions, and as a result the overall one, without requiring explicitly the accuracies of estimators in probability. In the second framework, local control barrier functions are constructed using the estimators' dynamics (without augmenting them with the subsystems' dynamics) in where we utilize a notion of so-called \emph{stochastic simulation functions} to compute a probabilistic bound on the estimation accuracy. We propose a sum-of-squares (SOS) optimization approach to search for local control barrier functions in both approaches, and accordingly, to compute the corresponding controllers. In order to illustrate the effectiveness of our proposed results, we apply both approaches to an adaptive cruise control problem.
	
	\section{Preliminaries and Problem Definition}
	\subsection{Preliminaries}
	A probability space in this work is presented by tuple $(\Omega,\mathcal{F}_{\Omega},\mathbb{P}_{\Omega})$, where $\Omega$ is a sample space, $\mathcal{F}_{\Omega}$ is a sigma-algebra on $\Omega$, and $\mathbb{P}_{\Omega}$ is a probability measure that assigns probabilities to events. Random variables introduced here are measurable functions of the form $X:(\Omega,\mathcal{F}_\Omega)\to (\mathcal{S}_X,\mathcal{F}_{\Omega})$ such that any random variable $X$ induces a probability measure on its space $(\mathcal{S}_X,\mathcal{F}_{\Omega})$ as \textit{Prob}$\{A\} = \mathbb{P}_\Omega\{X^{-1}(A)\}$ for any $A\in \mathcal{F}_X$. We directly present the probability measure on $(\mathcal{S}_X,\mathcal{F}_X)$ without explicitly mentioning the underlying probability space and the function $X$ itself.
	
	We call the topological space $\mathcal{S}$ as a Borel space if it is homeomorphic to a Borel subset of a Polish space. Euclidean space $\mathbb{R}^n$, its Borel subsets endowed with a subspace topology, and hybrid spaces are examples of Borel spaces. A Borel sigma-algebra is denoted by $\mathfrak{B}(\mathcal{S})$, where any Borel space $\mathcal{S}$ is assumed to be endowed with it. A map $f : \mathcal{S} \to Y$ is measurable whenever it is Borel measurable.
	
	\subsection{Notation}
	The sets of nonnegative and positive integers are denoted by $\mathbb{N}:=\{0,1,2,\ldots\}$ and $\mathbb{N}_{\geq 1}:= \{1,2,3,\ldots \}$, respectively. Moreover, symbols $\mathbb{R},\mathbb{R}_{>0}$, and $\mathbb{R}_{\geq 0}$ denote, respectively, the sets of real, positive and nonnegative real numbers. Given $N$ vectors $x_i\in \mathbb{R}^{n_i}, n_i\in \mathbb{N}_{\geq 1}$, $i\in \{1,\ldots, N\}$, we use $x=[x_1;\ldots;x_N]$ to denote the corresponding column vector of the dimension $\sum_{i}^{}n_i$. We denote by $\|\cdot\|$ the infinity norm. Given any $a\in \mathbb{R}$, $\lvert a \lvert$ denotes the absolute value of $a$. The identity function and composition of functions are denoted by $\mathcal{I}_d$ and the symbol $\circ $, respectively. A function $\kappa:\mathbb{R}_{\geq 0}\to \mathbb{R}_{\geq 0}$, is said to be a class $\mathcal{K}$ function if it is continuous, strictly increasing, and $\kappa(0)=0$. A class $\mathcal{K}$ function $\kappa(\cdot)$ is said to be a class $\mathcal{K}_\infty$ if $\kappa(r)\to\infty$ as $r \to\infty$. We denote the empty set by $\emptyset$. Given functions $f_i : X_i \to Y_i$, for any $i\in\{1,\ldots, N\}$, their Cartesian product $\prod ^{N}_{i=1}f_i: \prod^{N}_{i=1} X_i \to\prod ^{N}_{i=1} Y_i$ is defined as $(\prod^{N}_{i=1} f_i)(x_1,\ldots,x_N) = [f_1(x_1);\ldots;f_N(x_N)]$.
	
	\subsection{Partially-Observed Discrete-Time Stochastic Control Systems (a.k.a. Continuous-Space POMDPs)}
	In this paper, we consider partially-observed discrete-time stochastic control systems as formalized in the following definition.
	\begin{definition}\label{def_sigma_tuple}
		A partially-observed discrete-time stochastic control system $($PO-dt-SCS$)$ in this paper is characterized by the tuple
		\begin{equation}\label{eq:Sigma_tuple}
			\Sigma=(X,U,W,\varsigma_1,f,Y_1,Y_2,h_1,h_2,\varsigma_2),
		\end{equation}
		where,
		\begin{itemize}
			\item
			$X\subseteq \mathbb R^n$ is a Borel space as the state space of the
			system. The measurable space with $\mathfrak{B}(X)$ being the Borel sigma-algebra on the state space is denoted by $(X,\mathfrak{B}(X))$;
			\item $U\subseteq \mathbb R^m$ is a Borel space as the \emph{external} input space of the system; 
			\item $W\subseteq \mathbb R^p$ is a Borel space as the \emph{internal} input space of the system;
			\item $\varsigma_i$, $i\in\{1,2\}$, denote sequences of independent and identically distributed $($i.i.d.$)$ random variables from a sample space $\Omega$ to the set $\mathcal V_{\varsigma_i}$,
			\[
			\varsigma _i=\{\varsigma_i(k): \Omega \to \mathcal V_{\varsigma_i}, k \in \mathbb{N}\},
			\]
			\item
			$f:X\times U\times W\times \mathcal V_{\varsigma_1}\to X$ is a measurable function characterizing the state evolution of the system;
			\item 
			$Y_1 \subseteq \mathbb{R}^p$ is a Borel space as the  \textit{internal} output space of the system;
			\item
			$Y_2 \subseteq \mathbb{R}^q$ is a Borel space as the \textit{external} output space of the system;
			\item 
			$h_1:X\to Y_1$ is a measurable function that maps a state $x\in X$ to its \textit{internal} output $y_1=h_1(x)$;
			\item
			$h_2:X\times \mathcal{V}_{\varsigma_2}\to Y_2$ is a measurable function that maps a state $x(k)\in X$ to its \textit{external} output $y_2(k)=h_2(x(k),\varsigma_2(k))$.
			
		\end{itemize}
	\end{definition}
	
	An evolution of the state of PO-dt-SCS $\Sigma$ for a given initial state $x(0)\in X$ and input sequences $\upsilon(\cdot):\mathbb{N}\to U$ and $w(\cdot):\mathbb{N}\to W$ is described by	
	\begin{equation}\label{eq:Sigma}
		\Sigma:\begin{cases}
			x(k+1)=f(x(k),\upsilon(k),w(k),\varsigma_1(k)),\\
			y_1(k)=h_1(x(k)),\\
			y_2(k)=h_2(x(k),\varsigma_2(k)),\ \ \ \ \ \ \  \  \ \  \ k\in\mathbb{N}.\\
		\end{cases}
	\end{equation}
	
	A PO-dt-SCS $\Sigma$ in \eqref{eq:Sigma_tuple} can be \textit{equivalently} represented as a partially-observed Markov decision process $($POMDP$)$ \cite[Proposition 7.6]{kallenberg1997foundations}
	\begin{align}\label{MDP}
		\Sigma=(X,U,W,T_x,Y_1,Y_2,h_1,h_2,\varsigma_2),
	\end{align} where the map $T_{\mathsf x}:\mathfrak{B}(X)\times X\times U\times W\to [0,1]$ is a conditional stochastic kernel that assigns to any $x(k)\in X$, $\upsilon(k)\in U$, and $w(k)\in W$, a probability measure $T_{\mathsf x}(\cdot\mid x(k), \upsilon(k), w(k))$ on the measurable space $(X,\mathfrak{B}(X))$ so that for any set $\mathcal{A}\in\mathfrak{B}(X)$,
	\begin{align*}
		\mathbb{P}\big{(}x(k+1)\in\mathcal{A} \,\big |\, x(k),\upsilon(k),w(k))=
	 \int_{\mathcal{A}}T_{x}(x(k+1)\,\big |\, x(k),\upsilon(k),w(k)\big{)}.
	\end{align*}
	For given inputs $\upsilon(\cdot)$, and $w(\cdot)$, the stochastic kernel $T_{x}$ captures the evolution of the state of $\Sigma$ and can be uniquely determined by the pair $(\varsigma_1,f)$ from \eqref{eq:Sigma_tuple}. 
	Since two systems~\eqref{eq:Sigma_tuple} and~\eqref{MDP} are indeed equivalent, we interchangeably employ terms PO-dt-SCS and POMDP in the remainder of the paper.
	We associate to $U$ and $W$  sets $\mathcal{U}$ and $\mathcal{W}$, respectively, to be collections of sequences $\{\upsilon(k):\Omega\to U, k\in \mathbb{N}\}$ and $\{w(k):\Omega\to W, k\in \mathbb{N}\}$, in which $\upsilon (k)$ and $w(k)$ are independent of $\varsigma_i(l)$ for any $k,l\in\mathbb{N}$, $l\geq k$ and $i\in\{1,2\}$. The random sequences $x_{a\upsilon w}:\Omega\times \mathbb{N}\to X$, $y_{1_{a \upsilon w}}:\Omega\times \mathbb{N}\to Y_1$, and $y_{2_{a \upsilon w}}:\Omega\times \mathbb{N}\to Y_2$ satisfying \eqref{eq:Sigma} are called respectively the \textit{solution process}, \textit{internal output} and \textit{external output processes} of $\Sigma$, respectively, under an external input $\upsilon$, an internal input $w$, and an initial state $a$. 
	
	Since the main goal of this work is to study networks of  systems, the tuple representing interconnected systems, not containing internal inputs and outputs, is 
	$\Sigma=(X,U,\varsigma_1,f,Y,h,\varsigma_2)$, where $f:X\times U\times \mathcal V_{\varsigma_1}\to X$, and 
	\begin{equation}\label{Eq_2a} 
		\Sigma:\begin{cases}
			x(k+1)=f(x(k),\upsilon(k),\varsigma_1(k)),\\
			y(k)=h(x(k),\varsigma_2(k)),\ \ \  \ k\in\mathbb{N}.
		\end{cases}
	\end{equation}
	
	For the sake of controller synthesis using barrier certificates explained later in detail, we raise the following assumption on the existence of an estimator that estimates the state of the PO-dt-SCS in \eqref{eq:Sigma}.

	\begin{assumption}
		Consider a PO-dt-SCS $\Sigma=(X,U,W,\varsigma_1,f,Y_1,Y_2,h_1,h_2,\varsigma_2)$. States of $\Sigma$ in \eqref{eq:Sigma} can be estimated by a proper estimator  $\widehat{\Sigma}$ which is characterized by the tuple $\widehat{\Sigma}=(X,U,W, \hat{f}, Y_1,Y_2, h_1)$ and represented in the following form:
		\begin{equation}\begin{aligned}\label{eq:hat_Sigma_i}
				\widehat{\Sigma}:\begin{cases} \hat{x}(k+1)=\hat{f}(\hat{x}(k),\upsilon(k),\hat{w}(k),y_2(k)),\\
					\hat{y}_1(k) = h_1(\hat{x}(k)),\end{cases}
		\end{aligned}	\end{equation} 
		where $\upsilon$ and $y_2$ are external input and output signals of $\Sigma$ and $\hat{w}$ is the internal input signal coming from other estimators. We explain later how $\hat{w}$ is being fed by the estimators of other neighbouring subsystems.
		
		There exist numerous results in the relevant literature for the design of the estimator in \eqref{eq:hat_Sigma_i} for different classes of stochastic systems (cf. \cite{liang2009state, shen2011bounded, wang2013robust,stankovic2009consensus}). 
	\end{assumption}
	
	In the next section, we introduce notions of local control
	barrier functions (LCBF) and control barrier functions (CBF) for respectively POMDPs (with both internal and external inputs) and interconnected POMDPs (without internal inputs and outputs).
	
	\section{(Local) Control Barrier Functions}\label{CBD_section}
	First, we define (local) control barrier functions ((L)CBF) over an augmented system consisting of the stochastic (sub)system's and its estimator's dynamics. This formulation enables one to search for (local) control barrier functions with no prior knowledge of the estimation accuracy. Second, we formulate (local) control barrier functions over the estimator's dynamics (without augmenting them with the subsystem's dynamics) by utilizing a given probability bound on the estimation accuracy computed via a notion of so-called stochastic simulation functions. 
	
	\subsection{Notions of (L)CBF without considering  the estimation accuracy}\label{Barrier_1}
	Here, we first define the augmented process $\begin{bmatrix} x(k);\hat{x}(k)\end{bmatrix}$, where $x(k)$ and $\hat{x}(k)$ are the solution processes of subsystems  $\Sigma$ in \eqref{eq:Sigma} and their estimators $\widehat{\Sigma}$ in \eqref{eq:hat_Sigma_i}, respectively. The corresponding augmented stochastic  subsystem $\widetilde{\Sigma}$ can be defined as:
	\begin{equation}\label{eq:augmented_i}
		\widetilde{\Sigma}:
		\begin{bmatrix}x(k+1)\\ \hat{x}(k+1)\end{bmatrix}=
		\begin{bmatrix} f(x(k),\upsilon(k),w(k),\varsigma_{1}(k))\\\hat{f}(\hat{x}(k),\upsilon(k),\hat{w}(k),y_2(k)) \end{bmatrix}\!.
	\end{equation}
	
	Now, the local control barrier function is defined for system $\widetilde{\Sigma}$ in \eqref{eq:augmented_i}. This framework allows us to  provide one of our main results without any prior knowledge of the probabilistic distance between the actual states and their estimations. We now formally define local control barrier functions constructed over the augmented system $\widetilde{\Sigma}$.
	\begin{definition}\label{definition_barrier}
		Consider a POMDP $\Sigma$ in \eqref{eq:Sigma}, its estimator $\widehat{\Sigma}$ in \eqref{eq:hat_Sigma_i}, and
		the resulting augmented system $\widetilde{\Sigma}$ in \eqref{eq:augmented_i}. Let  $X_a, X_b\subseteq X$ represent some initial and unsafe regions, respectively. A function $\mathcal{B}: X\times X\rightarrow \mathbb R_{\geq 0}$ is called a local control barrier function $($LCBF$)$ for $\widetilde{\Sigma}$ if there exist constants $\bar{\psi},\bar{\gamma}\in\mathbb{R}_{\geq 0}$ and $\bar{\lambda}\in\mathbb{R}_{> 0}$, such that
		\begin{itemize}
			\item $\forall (x,\hat{x})\in X\times X$, \begin{equation}\label{eq:barrier_w_1}\mathcal{B}(x,\hat{x})\geq \alpha(\|\begin{bmatrix}h_1(x)\\h_1(\hat{x})\end{bmatrix}\|^{^2})
				,\end{equation}
		\end{itemize}
		\begin{itemize}
			\item $\forall (x,\hat{x})\in X_a\times X_a$, \begin{equation}\label{eq:barrier_w_2}\mathcal{B}(x,\hat{x})\leq \bar{\gamma},\end{equation}
		\end{itemize}
		\begin{itemize}
			\item $\forall (x,\hat{x})\in X_b\times X $, \begin{equation}\label{eq:barrier_w_3} \mathcal{B}(x,\hat{x})\geq \bar{\lambda},\end{equation}
		\end{itemize}
		\begin{itemize}
			\item 	$\forall \hat{x}(k)\in X$, $\forall \hat{w}(k)\in W$, $\exists \upsilon(k)\in U$, such that $\forall x(k)\in X$, $\forall w(k)\in W$, 
		\end{itemize}

		\begin{equation}\begin{aligned}\label{eq:barrier_w_4} 
				\mathbb{E} \Big{[}&\mathcal{B} \big{(}f(x(k),\upsilon(k),w(k),\varsigma_{1}(k)), \hat{f}(\hat{x}(k),\upsilon(k),\hat{w}(k),y_2(k)) \big{)}
					 \,\big|\, x(k), \hat{x}(k), \upsilon(k), w(k), \hat{w}(k) \Big{]}
				\\&	\leq \max \Big{\{}\bar{\kappa}\mathcal{B}(x(k),\hat{x}(k)),\rho(\|\begin{bmatrix}w(k)\\ \hat{w}(k)\end{bmatrix}\|^{^2}),\bar{\psi}\Big{\}},
		\end{aligned}\end{equation}
		for some $\alpha \in \mathcal{K}_\infty$, $\rho \in \mathcal{K}_\infty\cup \{0\}$ and $0<\bar{\kappa} < 1$.
	\end{definition}
	Definition \ref{definition_barrier} can also be stated for interconnected systems without internal inputs and outputs by eliminating all the terms related to the internal input $w$, its estimation $\hat{w}$, internal output $h_1(x)$, and its estimation $h_1(\hat x)$ as defined below.

	\begin{definition}\label{definition_barrier_no_w}
		Consider an (interconnected) POMDP  $\Sigma=(X,U,\varsigma_1,f,Y,h,\varsigma_2)$, its estimator $\widehat{\Sigma}$ also
		without internal inputs and outputs, and the augmented system $\widetilde{\Sigma}=[\Sigma ; \widehat{\Sigma}]$. Let  $X_a, X_b\subseteq X $,  respectively, represent initial and unsafe regions. A function $\mathcal{B}: X\times X\rightarrow \mathbb R_{\geq 0}$ is called a control barrier function (CBF) for $\widetilde{\Sigma}$ if there exist constants $\psi,\gamma\in\mathbb{R}_{\geq 0}$ and $\lambda\in\mathbb{R}_{> 0}$ such that $\gamma < \lambda $, and
		\begin{itemize}
			\item $\forall (x,\hat{x})\in X_a\times X_a$, \begin{equation}\label{eq:barrier_no_w_1}\mathcal{B}(x,\hat{x})\leq \gamma,\end{equation}
		\end{itemize}
		\begin{itemize}
			\item $\forall (x,\hat{x})\in X_b \times X $, \begin{equation} \label{eq:barrier_no_w_2}
				\mathcal{B}(x,\hat{x})\geq \lambda,
			\end{equation}
		\end{itemize}
		\begin{itemize}
			\item and $\forall \hat{x}(k)\in X$, $\exists \upsilon(k)\in U$, such that $\forall x(k)\in X$,
		\end{itemize}

		\begin{equation}\begin{aligned}\label{eq:barrier_no_w_3}
				\mathbb{E}\Big{[}&\mathcal{B}\big{(}f(x(k),\upsilon(k),\varsigma_{1}(k)), \hat{f}(\hat{x}(k),\upsilon(k),y(k))\big{)}\,\big|\, x(k), \hat{x}(k), \upsilon(k)
				\Big{]}
				\leq \max \big{\{}\kappa\mathcal{B}(x(k),\hat{x}(k)),\psi\big{\}},
			\end{aligned}
		\end{equation}
		for some $0 <\kappa < 1$.
	\end{definition}
	\begin{remark}
		Note that we need the condition $\gamma < \lambda$ (\emph{i.e.,} $X_a\cap X_b=\emptyset$) in order to provide a meaningful probability in Theorem \ref{theorem_bound} later. This requirement is only for the interconnected system and not for subsystems. In particular, LCBFs are mainly utilized for the compositional construction of CBFs over interconnected systems and are not directly employed for ensuring the probability of safety satisfaction. The above definition associates a policy $\eta:X\to U$ to a CBF, where $X$ here is the state set of the estimator $\widehat\Sigma$. Definition \ref{definition_barrier_no_w} gives such a policy according to the existential quantifier over the input for any estimator's state $\hat x\in X$.
	\end{remark}
	The next theorem shows the usefulness of having a CBF to quantify an upper bound on the exit probability $($\textit{i.e.,} the probability that the solution process of the interconnected system reaches the unsafe region in a finite-time horizon$)$ of POMDP (without  internal inputs and outputs).
	\begin{theorem}\label{theorem_bound}
		Let $\Sigma=(X,U,\varsigma_1,f,Y,h,\varsigma_2)$ be a POMDP (without internal inputs and outputs) and $\widehat{\Sigma}$ be its corresponding estimator. Suppose $\mathcal{B}$ is a CBF according to Definition \ref{definition_barrier_no_w}. Then, the probability that the solution process of $\Sigma$ starts from any initial states $x(0)=a \in X_a$ and reaches $X_b$ under the control policy $\eta$ within a time horizon $[0,T_d]$ is formally upper bounded as 
		\begin{equation}\label{eq:delta_bound}\begin{aligned}
				\mathbb{P}&\Big{[} x_{a\upsilon}(k)\in X_b \ \text{for some} \ k \in [0, T_d]
				\ 	\big |  \ a,\upsilon\Big{]}
				\leq \delta,\end{aligned}\end{equation}
		where,
		\begin{equation}\begin{aligned}\label{delta}
				\delta:=\begin{cases}
					1-(1-\frac{\gamma}{\lambda})(1-\frac{\psi}{\lambda})^{T_d}, &\text{if } \ \lambda\geq \frac{\psi}{\kappa},
					\\
					\\\frac{\gamma}{\lambda}(1-\kappa)^{T_d}+(\frac{\psi}{\kappa\lambda})(1-(1-\kappa)^{T_d}), & \text{if } \ \lambda < \frac{\psi}{\kappa}.
				\end{cases}
		\end{aligned} \end{equation}
	\end{theorem}
	The proof of Theorem~\ref{theorem_bound} is provided in Appendix.
	\begin{remark}
		Utilizing the augmented system $\widetilde{\Sigma}$ as in \eqref{eq:augmented_i} provides us with the results in Theorem \ref{theorem_bound} with no need of knowing the estimation accuracy explicitly. This allows more flexibility in designing the estimator and potentially results in tighter upper bounds.
	\end{remark}
	
	In the next subsection, we formulate control barrier functions only over the estimators' dynamics by utilizing a probability bound on the estimation accuracy. 
	
	\subsection{Notions of (L)CBF by considering the estimation accuracy}\label{Barrier_2}
	Given an estimator with a probabilistic guarantee on the accuracy of the estimation,  we propose an approach to construct a CBF defined only over the states of the estimator $\widehat{\Sigma}$. For a given time horizon $T_d$, we assume the probabilistic bound on the accuracy of the estimator is given by \cite{reif2000stochastic}:
	\begin{equation}\begin{aligned}\notag 
			\forall  \epsilon >0, \exists \theta\in(0,1],  \text{\ such that}
			\quad \mathbb{P}\Big{[}\sup_{0\leq k \leq T_d}\|x_{a\upsilon }(k)-\hat{x}_{\hat{a} \upsilon }(k)\| < \epsilon \,\big|\, a,\hat{a}, \upsilon\Big{]}\geq 1-\theta,
	\end{aligned}\end{equation}
	for any $a,\hat a\in X$ and any $\upsilon\in\mathcal{U}$.
	In order to quantify the distance (a.k.a. error) between a system's state and its estimation, we employ notions of so-called stochastic (pseudo)-simulation functions. To do so, we first introduce stochastic pseudo-simulation functions (SPSF) for POMDPs with both internal and external inputs. We then define stochastic simulation functions (SSF) for interconnected POMDPs without internal inputs and outputs.
	\begin{definition}\label{def:phi_def_w}
		Consider a POMDP $\Sigma$ in \eqref{eq:Sigma} and its corresponding estimator $\widehat{\Sigma}$ in \eqref{eq:hat_Sigma_i}. A function $\mbox{\boldmath$\phi$}:X\times X\to\mathbb{R}_{\geq 0}$ is called a \textit{stochastic pseudo-simulation function} $($SPSF$)$ from $\widehat{\Sigma}$ to $\Sigma$ if
		\begin{enumerate}[(i)]
			\item $ 
			\forall x\in X, \forall \hat{x}\in X,$ $$\varepsilon (\|x-\hat{x}\|) \leq \mbox{\boldmath$\phi$}(x,\hat{x}),$$	
			\item   $\forall \hat{x}(k)\in X, \forall \hat{w}(k)\in W, \exists \upsilon(k)\in {U}$, such that $\forall x(k)\in X, \forall w(k)\in W$, 
			\begin{equation*}\begin{aligned}
					\mathbb{E}\Big{[}&\mbox{\boldmath{$\phi$}}\big{(}
					f(x(k),\upsilon(k),w(k),\varsigma_{1}(k)), \hat{f}(\hat{x}(k),\upsilon(k),\hat{w}(k),y_2(k)) \big{)}
						 \,\big|\, x(k), \hat{x}(k),\upsilon(k), w(k), \hat{w}(k)
					\Big{]}
					\\&	\leq \max\big{\{}\bar{\mu}\mbox{\boldmath{$\phi$}}(x(k),\hat{x}(k)),\varrho(\| w(k)-\hat{w} (k) \|),\bar{c}\big{\}},
			\end{aligned}\end{equation*}
		\end{enumerate}
		for some $0<\bar{\mu}<1$, ${\varepsilon} \in \mathcal{K}_\infty$,  $\varrho\in\mathcal{K}_\infty\cup \{0\}$, and $\bar{c}\in\mathbb{R}_{\geq 0}$.
	\end{definition}
	Definition \ref{def:phi_def_w} can also be stated for POMDPs without internal inputs and outputs by eliminating all the terms related to the internal input $w$ and its estimation $\hat{w}$ as defined below.
	\begin{definition}\label{def:phi_def}
		Consider an (interconnected) POMDP $\Sigma=(X,U,\varsigma_1,f,Y,h,\varsigma_2)$ and its estimator $\widehat{\Sigma}$. A function $\mbox{\boldmath$\phi$}:X\times X\to\mathbb{R}_{\geq 0}$ is called a \textit{stochastic simulation function} $($SSF$)$ from $\widehat{\Sigma}$ to $\Sigma$ if
		\begin{enumerate}[(i)]
			\item $\forall x\in X, \forall \hat{x}\in X,$  $$\varepsilon (\|x-\hat{x}\|) \leq \mbox{\boldmath$\phi$}(x,\hat{x}),$$	
			\item  $\forall\hat{x}(k)\in X, \exists \upsilon(k) \in {U}$, such that $\forall x(k)\in X$,  
			\begin{equation*}\begin{aligned}
					\mathbb{E}\Big{[}&\mbox{\boldmath{$\phi$}}\big{(}
					f(x(k),\upsilon(k),\varsigma_{1}(k)), \hat{f}(\hat{x}(k),\upsilon(k),y(k)) \big{)}
					\,
					\big|\, x(k), \hat{x}(k), \upsilon(k)\Big{]}
					\leq \max\big{\{}\mu\mbox{\boldmath{$\phi$}}(x(k),\hat{x}(k)),c\big{\}},
			\end{aligned}\end{equation*}
		\end{enumerate}
		for some $0<\mu<1$, $\varepsilon \in \mathcal{K}_\infty$, and $c\in\mathbb{R}_{\geq 0}$.
	\end{definition}
	The next theorem shows how an SSF can be employed to obtain the probability bound on the estimation accuracy.
	\begin{theorem}\label{thm:SSF}
		Consider a POMDP $\Sigma$ in \eqref{Eq_2a}, its estimator $\widehat{\Sigma}$ in \eqref{eq:hat_Sigma_i} (without internal inputs and outputs), and $\epsilon >0$. Suppose $\mbox{\boldmath{$\phi$}}$ is an SSF from $\widehat{\Sigma}$ to $\Sigma$. For any $\upsilon\in\mathcal{U}$,
		and for any random variables $a$ and $\hat{a}$ as initial states of $\Sigma$ and $\widehat{\Sigma}$, respectively, the following inequality holds:
		\[ \mathbb{P}\Big{[}\sup_{0\leq k \leq T_d}\|x_{a\upsilon }(k)-\hat{x}_{\hat{a} \upsilon}(k)\|\geq \epsilon\,\big|\, a,\hat{a}, \upsilon \Big{]}\leq \theta,\]
		where,
		\begin{equation}\label{eq:xhat_bound}
			\theta:=\begin{cases}
				1-(1-\frac{\mbox{\boldmath{$\phi$}}(a,\hat{a})}{\varepsilon(\epsilon)})(1-\frac{c}{\varepsilon(\epsilon)})^{T_d},
				\ \ \ \ \ \text{\small if} \ \varepsilon(\epsilon)\geq \frac{c}{\mu},\\
				\\
				(\frac{\mbox{\boldmath{$\phi$}}(a,\hat{a})}{\varepsilon(\epsilon)})(1-\mu)^{T_d}+(\frac{c}{\mu\varepsilon(\epsilon)})(1-(1-\mu)^{T_d}),\\ \ \ \ \ \ \ \ \ \ \ \ \ \  \ \ \ \ \ \ \ \ \ \ \ \ \ \ \ \ \ \ \ \ \ \ \ \ \ \ \ \text{\small if}\ \varepsilon(\epsilon)<\frac{c}{\mu}.
			\end{cases}
		\end{equation} 
	\end{theorem}
	
	The proof of Theorem~\ref{thm:SSF} is provided in Appendix.
	
	We now propose our second formulation of control barrier functions defined only over the estimators' dynamics as the following. 
	
	\begin{definition}\label{definition_barrier_xhat}
		Consider a POMDP $\Sigma$ as in \eqref{eq:Sigma}, its estimator $\widehat{\Sigma}$, and $\epsilon >0$. Let $X_{a} , X_{b}\subseteq X$ denote respectively initial and unsafe sets. Let us define $X_b^\epsilon:=\{\hat{x}\in X\mid \exists x\in X_b,\|\hat{x}-x\|\leq \epsilon\}$ (\emph{i.e.,} unsafe set for  $\widehat{\Sigma}$). A function $\mathcal{B}: X \rightarrow \mathbb R_{\geq 0}$ is called a local control barrier function $($LCBF$)$ for ${\widehat{\Sigma}}$ if there exist constants $\bar{\psi},\bar{\gamma}\in\mathbb{R}_{\geq 0}$ and $\bar{\lambda}\in\mathbb{R}_{> 0}$, such that
		\begin{itemize}
			\item $\forall {x}\in X$, \begin{equation}\label{eq:barrier_hat_w_1}\mathcal{B}({x})\geq \alpha(\|h_1(x)\|^{^2}),\end{equation}
		\end{itemize}
		\begin{itemize}
			\item $\forall {x}\in X_a$, \begin{equation}\label{eq:barrier_hat_w_2}\mathcal{B}({x})\leq \bar{\gamma},\end{equation}
		\end{itemize}
		\begin{itemize}
			\item 
			$\forall {x}\in X_b^{\epsilon} $, \begin{equation}\label{eq:barrier_hat_w_3} \mathcal{B}(x)\geq \bar{\lambda},\end{equation}
		\end{itemize}
		\begin{itemize}
			\item and $\forall \hat{x}(k)\in X$, $\forall \hat{w}(k)\in W$, $\exists \upsilon(k)\in U$, such that $\forall y_2(k)\in Y_2$,
		\end{itemize}
		\begin{equation}
			\begin{aligned}\label{eq:barrier_hat_w_4} 
				\mathbb{E}\Big{[}&\mathcal{B}\big{(}
				\hat{f}(\hat{x}(k),\upsilon(k),\hat{w}(k),y_2(k))
				\big{)}\mid \hat{x}(k), \upsilon(k), \hat{w}(k) \Big{]}
				\leq \max \big{\{}\bar{\kappa}\mathcal{B}(\hat{x}(k)),\rho( \|\hat{w}(k)\|^{^2} ),\bar{\psi}\big{\}},
			\end{aligned}	
		\end{equation}
		for some $0<\bar{\kappa} < 1$, $\alpha \in \mathcal{K}_\infty$, and $\rho \in \mathcal{K}_\infty\cup \{0\}$.
	\end{definition}
	
	We now modify Definition \ref{definition_barrier_xhat} and present it for the interconnected POMDPs as the following.
	\begin{definition}\label{definition_barrier_no_w_xhat}
		Consider an (interconnected) POMDP $\Sigma=(X,U,\varsigma_1,f,Y,h,\varsigma_2)$, its estimator $\widehat{\Sigma}$ without internal inputs and outputs and $\epsilon>0$. Let $X_a, X_{b}\subseteq X$ denote respectively initial and unsafe sets. Let us define $X_b^\epsilon:=\{\hat{x}\in X\mid \exists x\in X_b,\|\hat{x}-x\|\leq \epsilon\}$. A function $\mathcal{B}: X\rightarrow \mathbb R_{\geq 0}$ is called a control barrier function for $\widehat{\Sigma}$ if there exist constants $\psi,\gamma\in\mathbb{R}_{\geq 0}$ and $\lambda\in\mathbb{R}_{> 0}$ such that $\gamma < \lambda $ and
		\begin{itemize}
			\item $\forall {x}\in X_a$, \begin{equation*}
				\mathcal{B}({x})\leq \gamma,\end{equation*}
		\end{itemize}
		\begin{itemize}
			\item $\forall {x}\in X_b^\epsilon $, \begin{equation*}
				\mathcal{B}({x})\geq \lambda,
			\end{equation*}
		\end{itemize}
		\begin{itemize}
			\item and $\forall \hat{x}(k)\in X$, $\exists \upsilon(k)\in U$, such that $\forall y(k)\in Y$,
		\end{itemize}

		\begin{equation*}\begin{aligned}
				\mathbb{E}\Big{[}\mathcal{B}\big{(}
				\hat{f}(\hat{x}(k),\upsilon(k),y(k))
				\big{)} \mid \hat{x}(k), \upsilon(k) \Big{]}\leq
				\max \big{\{}\kappa\mathcal{B}(\hat{x}(k)),\psi\big{\}},
		\end{aligned}\end{equation*}
		for some $0 <\kappa < 1$.
	\end{definition}

	One can employ Definition \ref{definition_barrier_no_w_xhat} and provide a similar result as Theorem \ref{theorem_bound}:
	\begin{equation*}
		\mathbb{P}\Big{[} \hat{x}_{\hat{a}\upsilon}(k)\in X_b^{\epsilon}\ \text{for some} \  k \in [0,T_d] \ \big|\ \hat{a},\upsilon\Big{]}\leq \delta,
	\end{equation*}
	where $\delta$ is computed as in~\eqref{delta}. In the next Theorem, we provide an upper bound on the exit probability of POMDP using the estimation accuracy.

	\begin{theorem}\label{theorem_bound_xhat}
		Let $\Sigma=(X,U,\varsigma_1,f, Y, h,\varsigma_2)$ be a POMDP without internal inputs and outputs, $\widehat{\Sigma}$ be its corresponding estimator and $\epsilon$ be a positive constant. Suppose $\mathcal{B}$ is a CBF for $\widehat{\Sigma}$ as in Definition \ref{definition_barrier_no_w_xhat}. Then, the probability that the solution process of $\Sigma$ starts from any initial state $x(0)=a\in X_a$ and reaches $X_b$ under control policy $\eta$ within a time horizon $[0,T_d]$ is upper bounded as
		\begin{equation}
			\mathbb{P}\Big{[} x_{a\upsilon}(k)\in X_b\ \text{for some} k \in [0,T_d] \ \big|\ a,\upsilon \Big{]}
			\leq \delta+\theta,
		\end{equation}
		where $\delta$ and $\theta$ are computed as in \eqref{delta} and \eqref{eq:xhat_bound}, respectively.
	\end{theorem}
	
	The proof is similar to that of \cite[Theorem 3.3]{jahanshahi2020synthesis} and is omitted here due to lack of space.
	
	\begin{remark}
		Note that the first proposed approach does not require a prior knowledge of the estimation accuracy, and accordingly, it gives the user more flexibility on the estimator design. Moreover, in the first approach the computation of the exit probability can be done in one shot without utilizing SSFs and, hence, be less conservative. However, the computational complexity in the first approach is more than the second one since the control barrier function should be constructed over the augmented system.
	\end{remark}
	
	In the next sections, we analyze networks of POMDP and discuss under which conditions one can construct a CBF of an interconnected system based on LCBF of its subsystems.

	\section{Interconnected POMDP}
	We consider a collection of partially-observed stochastic control subsystems and their estimators as
	\begin{equation*}\begin{aligned}
			\Sigma_i&=(X_i,U_i,W_i,\varsigma_{1i},f_i,Y_{1_i},Y_{2_i},h_{1_i},h_{2_i},\varsigma_{2_i}),\\
			\widehat{\Sigma}_i&=(X_i,U_i,W_i,\hat{f}_i,Y_{1_i},Y_{2_i},h_{1_i}),  \ i\in\{1,\ldots, N\},
	\end{aligned}\end{equation*}
	
	where internal inputs and outputs are partitioned as 
	\begin{equation}\begin{aligned}\label{eq:wi_and_yi}
			w_i&=\begin{bmatrix}w_{i1}; \ldots; w_{i(i-1)};  w_{i(i+1)};  \ldots;  w_{iN}\end{bmatrix}\!,\\
			y_{1_i}&=\begin{bmatrix}y_{1_{i1}}; \ldots   y_{1_{i(i-1)}};  y_{1_{i(i+1)}};  \ldots;  y_{1_{iN}}\end{bmatrix}\!,\\
	\end{aligned}\end{equation}
	and their internal output spaces and functions are of the form 
	\begin{equation}\begin{aligned}\label{eq:Y_and_h_1}
			Y_{1_i}&=\prod_{j=1,j\neq i}^NY_{1_{ij}}, 
			\\h_{1_i}(x_i)&=[h_{1_{i1}}(x_i);\ldots ;h_{1_{i(i-1)}}(x_i);h_{1_{i(i+1)}}(x_i);\ldots h_{1_{iN}}(x_i)].
	\end{aligned}\end{equation}
	Furthermore, the internal input and output of the estimators are also partitioned similar to \eqref{eq:wi_and_yi} and \eqref{eq:Y_and_h_1}.
	
	Outputs $y_{1_{ij}}$ with $i\neq j$ are \textit{internal} outputs which are employed for the sake of interconnections. If there is a connection from $\Sigma_j$ to $\Sigma_i$, we assume that $w_{ij}$ is equal to $y_{1_{ji}}$. Otherwise,  the connecting output function is identically zero, \textit{i.e.,} $h_{1_{ji}}\equiv 0$. The same interconnections hold for the estimators. If there is a connection from $\widehat{\Sigma}_j$ to $\widehat{\Sigma}_i$, we assume that $\hat{w}_{ij}$ is equal to $\hat{y}_{1_{ji}}$. Otherwise, the connecting output function is identically zero, \textit{i.e.,} $h_{1_{ji}}\equiv 0$. Now we define interconnected partially-observed stochastic control systems.
	\begin{definition}\label{def_sub_tule}
		Consider $N\in\mathbb{N}_{\geq 1}$ POMDPs $\Sigma_i=(X_i,U_i,W_i,\varsigma_{1_i},f_i,Y_{1_i},Y_{2_i},h_{1_i},h_{2_i},\varsigma_{2_i})$, $i\in\{1,\ldots,N\}$, with the input-output configuration as in \eqref{eq:wi_and_yi}-\eqref{eq:Y_and_h_1}. The interconnection of $\Sigma_i$, for any $i\in\{1,\ldots N\}$, is the interconnected POMDP $\Sigma=(X,U,\varsigma_1,f,Y,h,\varsigma_2)$, denoted by $\mathcal I(\Sigma_1,\ldots,\Sigma_N)$, such that $X:=\prod_{i=1}^N X_i$, $U:=\prod_{i=1}^N U_i$, $\varsigma_1=\begin{bmatrix}\varsigma_{1_1}; \cdots; \varsigma_{1_N} \end{bmatrix}$, $f:=\prod_{i=1}^Nf_i$, $Y:=\prod_{i=1}^N Y_{i}$, $h:=\prod_{i=1}^Nh_{i}$,  and $\varsigma_2=\begin{bmatrix}\varsigma_{2_1}; \cdots; \varsigma_{2_N} \end{bmatrix}$, subjected to the following constraint:
		\begin{equation*}
			\forall i,j\in\{1,\ldots,N\},i\neq j: \ \ \ w_{ji}=y_{1_{ij}}, \ \ \ Y_{1_{ij}}\subseteq W_{ji}.
		\end{equation*}
	\end{definition}
	
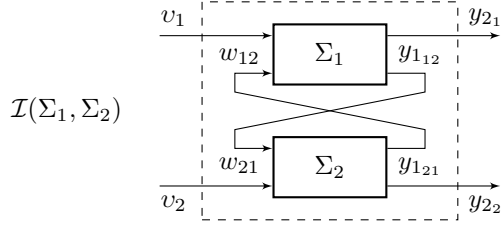
\begin{figure}[ht]
		\begin{tikzpicture}[>=latex']
			\tikzstyle{block} = [draw, 
			thick,
			rectangle, 
			minimum height=.8cm, 
			minimum width=1.5cm]
			\node at (-3.5,-0.75) {$\mathcal{I}(\Sigma_1,\Sigma_2)$};
			\draw[dashed] (-1.7,-2.2) rectangle (1.7,.7);
			\node[block] (S1) at (0,0) {$\Sigma_1$};
			\node[block] (S2) at (0,-1.5) {$\Sigma_2$};
			\draw[->] ($(S1.east)+(0,0.25)$) -- node[very near end,above] {$y_{2_1}$} ($(S1.east)+(1.5,.25)$);
			\draw[<-] ($(S1.west)+(0,0.25)$) -- node[very near end,above] {$\upsilon_{1}$} ($(S1.west)+(-1.5,.25)$);
			\draw[->] ($(S2.east)+(0,-.25)$) -- node[very near end,below] {$y_{2_2}$} ($(S2.east)+(1.5,-.25)$);
			\draw[<-] ($(S2.west)+(0,-.25)$) -- node[very near end,below] {$\upsilon_{2}$} ($(S2.west)+(-1.5,-.25)$);
			\draw[->] 
			($(S1.east)+(0,-.25)$) -- node[very near end,above] {$y_{1_{12}}$} 
			($(S1.east)+(.5,-.25)$) --
			($(S1.east)+(.5,-.5)$) --
			($(S2.west)+(-.5,.5)$) --
			($(S2.west)+(-.5,.25)$) -- node[very near start,below] {$w_{21}$}
			($(S2.west)+(0,.25)$) ;
			\draw[->] 
			($(S2.east)+(0,.25)$) -- node[very near end,below] {$y_{1_{21}}$} 
			($(S2.east)+(.5,.25)$) --
			($(S2.east)+(.5,.5)$) --
			($(S1.west)+(-.5,-.5)$) --
			($(S1.west)+(-.5,-.25)$) -- node[very near start,above] {$w_{12}$}
			($(S1.west)+(0,-.25)$) ;
		\end{tikzpicture}
		\caption{Interconnection of two POMDPs $\Sigma_1$ and $\Sigma_2$.}
		\label{fig:interconnection}
	\end{figure}

	In a similar way, we define the interconnection of estimators $\widehat{\Sigma}$ as the following.
	\begin{definition}\label{def_sub_tule1}
		Consider $N\in\mathbb{N}_{\geq 1}$ estimators $\widehat{\Sigma}_i=(X_i,U_i,W_i,\hat{f}_i,Y_{1_i},Y_{2_i},h_{1_i})$, $i\in\{1,\ldots,N\}$, with the input-output configuration similar to  \eqref{eq:wi_and_yi}-\eqref{eq:Y_and_h_1}. The interconnection of $\widehat{\Sigma}_i$, for any $i\in\{1,\ldots N\}$, is the interconnected estimator $\widehat{\Sigma}=(X,U,\hat{f},Y)$, denoted by $\mathcal I(\widehat{\Sigma}_1,\ldots,\widehat{\Sigma}_N)$, such that $X:=\prod_{i=1}^N X_i$, $U:=\prod_{i=1}^N U_i$, $\hat{f}:=\prod_{i=1}^N\hat{f}_i$, and $Y:=\prod_{i=1}^N Y_{i}$, subject to the following constraint:
		\begin{equation*}
			\forall i,j\in\{1,\ldots,N\},i\neq j: \ \ \ \hat{w}_{ji}=\hat{y}_{1_{ij}}, \ \ \ Y_{1_{ij}}\subseteq W_{ji}.
		\end{equation*}
	\end{definition}
	An example of the interconnection of two POMDPs $\Sigma_1$ and $\Sigma_2$ is illustrated in Fig.~\ref{fig:interconnection}.
	\section{Compositional Construction of CBF}
	In this section, we analyze networks of POMDP and provide a compositional approach to construct a CBF of an interconnected POMDP based on LCBF of its subsystems. For $i\in\{1,\ldots,N\}$, consider the PO-dt-SCS $\Sigma_i$ in \eqref{eq:Sigma}, its corresponding estimator $\widehat{\Sigma}_i$ in \eqref{eq:hat_Sigma_i}, and the augmented system $\widetilde{\Sigma}$ in \eqref{eq:augmented_i}. Assume there exists a LCBF $\mathcal{B}_i$ as defined in Definition \ref{definition_barrier} or \ref{definition_barrier_xhat} with functions $\alpha_i\in\mathcal{K}_\infty$, $\rho_i\in\mathcal{K}_\infty\cup\{0\}$ and constants $\bar{\lambda}_i, \bar{\psi}_i\in\mathbb{R}_{\geq 0}$, $\bar{\gamma}_i\in\mathbb{R}_{>0}$, and $0<\bar{\kappa}_i<1$. Now we raise the following small-gain assumption that is essential for the compositionality results of this section.

	\begin{assumption}\label{small_gain_assumption}
		Assume that $\mathcal{K}_\infty$ functions $\bar{\kappa}_{ij}$ defined as 
		\begin{equation*}
			\bar{\kappa}_{ij}(s):=\begin{cases}
				\bar{\kappa}_i(s),\ \ \ \ \ \ \ \ \ \ \ &\text{\small if } \ i=j,
				\\\rho_i(\alpha_j^{-1}(s)), \ \ \ \ \ \ \ \  \ \ \ \ \ \ \ \ \ & \text{\small if } \ i\neq j,
			\end{cases} 
		\end{equation*}
		satisfy
		\begin{equation}\label{eq:k_ij_o}
			\bar{\kappa}_{i_1 i_2}\circ \bar{\kappa}_{i_2 i_3}\circ \cdots \circ\bar{\kappa}_{i_{r-1} i_r}\circ \bar{\kappa}_{i_r i_1}<\mathcal{I}_d,
		\end{equation}
		for all sequences $(i_1,\ldots,i_r)\in \{1,\ldots,N\}^r$ and $r\in\{1,\ldots,N\}$.
	\end{assumption}

	\begin{remark}
		Note that the small-gain condition \eqref{eq:k_ij_o} is a standard one in studying the stability of large-scale interconnected systems via ISS Lyapunov functions \cite{dashkovskiy2007iss, dashkovskiy2010small}. This condition is automatically satisfied if each $\bar{\kappa}_{ij}$ is less than identity $(\bar{\kappa}_{ij} < \mathcal{I}_d, \forall i,j \in \{1,\ldots, N \})$.
	\end{remark}

	The small-gain condition \eqref{eq:k_ij_o} implies the existence of $\mathcal{K}_\infty$ functions $\sigma_i > 0$  \cite[Theorem 5.5]{ruffer2010monotone}, satisfying
	\begin{equation}\label{eq:sigma_k_ij}
		\max_{i,j}\big{\{}\sigma_i^{-1}\circ\bar{\kappa}_{ij}\circ\sigma_j\big{\}}<\mathcal{I}_d, \ \ \ \ \ \ \ i,j=\{1,\ldots,N\}.
	\end{equation}

	In the next theorem, we show that if Assumption \ref{small_gain_assumption} holds and $\max_{i} \sigma_i^{-1}$ is concave (in order to employ Jensen’s inequality), then one can compute a CBF for the interconnected system $\Sigma$ as in Definition \ref{definition_barrier_no_w} in a compositional fashion.
	\begin{theorem}\label{small_gain_theorem}
		Consider the interconnected POMDP $\Sigma=\mathcal{I}(\Sigma_i,\ldots,\Sigma_N)$ induced by $N\in\mathbb{N}_{\geq 1}$  subsystems $\Sigma_i$. Suppose that for each $\Sigma_i$ there exits an estimator $\widehat{\Sigma}_i$ together with a corresponding LCBF $\mathcal{B}_i$ as defined in Definition \ref{definition_barrier} with initial and unsafe sets $X_{a_i}$ and $X_{b_i}$, respectively. If Assumption \ref{small_gain_assumption} holds and $\max_i\sigma_i^{-1}$ for $\sigma_i$ as in \eqref{eq:sigma_k_ij} is concave and
		\begin{equation}\label{eq:gamma_einq_lambda}
			\max_i\{\sigma_i^{-1}(\bar{\gamma_i})\}< \max_i\{\sigma_i^{-1}(\bar{\lambda}_i)\},
		\end{equation}
		then function $\mathcal{B}(x,\hat{x})$ defined as
		\begin{equation}\label{eq:max_barrier}
			\mathcal{B}(x,\hat{x}):=\max_i \big{\{}\sigma_i^{-1}(\mathcal{B}_i(x_i,\hat{x}_i))\big{\}},
		\end{equation}
		is a CBF for the augmented system $\widetilde{\Sigma}=[\Sigma \ ; \ \widehat{\Sigma}]$ with initial and unsafe sets $X_a=\prod^{N}_{i=1}X_{a_i}$, $X_b=\prod^{N}_{i=1}X_{b_i}$, respectively.
	\end{theorem}
	
	The proof of Theorem \ref{small_gain_theorem} is provided in Appendix.
	
	Similarly, we propose the next theorem to compute a CBF for an interconnected system $\Sigma$ as in Definition \ref{definition_barrier_no_w_xhat} in a compositional way based on LCBFs of subsystems.
	\begin{theorem}\label{small_gain_theorem_2}
		Consider an interconnected POMDP $\Sigma=\mathcal{I}(\Sigma_i,\ldots,\Sigma_N)$ induced by $N\in\mathbb{N}_{\geq 1}$  subsystems $\Sigma_i$. Suppose that for each $\Sigma_i$ there exits an estimator $\widehat{\Sigma}_i$ together with a corresponding LCBF $\mathcal{B}_i$ as defined in Definition \ref{definition_barrier_xhat} with initial and unsafe sets $X_{a_i}$ and $X_{b_i}^\epsilon$, respectively. If Assumption \ref{small_gain_assumption} holds and $\max_i\sigma_i^{-1}$ for $\sigma_i$ as in \eqref{eq:sigma_k_ij} is concave and
		\begin{equation}\label{eq:gamma_einq_lambda1}
			\max_i\{\sigma_i^{-1}(\bar{\gamma_i})\}< \max_i\{\sigma_i^{-1}(\bar{\lambda}_i)\},
		\end{equation}
		then function $\mathcal{B}(x)$ defined as
		\begin{equation}\label{eq:max_barrier1}
			\mathcal{B}(x):=\max_i \big{\{}\sigma_i^{-1}(\mathcal{B}_i(x_i))\big{\}},
		\end{equation}
		is a CBF for the estimator $\widehat{\Sigma}=\mathcal{I}(\widehat{\Sigma}_i,\ldots,\widehat{\Sigma}_N)$ with initial and unsafe sets $X_a=\prod^{N}_{i=1}X_{a_i}$, $X_b^\epsilon=\prod^{N}_{i=1}X_{b_i}^\epsilon$, respectively. 
	\end{theorem}
	The proof of Theorem \ref{small_gain_theorem_2} follows the same reasoning as that of  Theorem \ref{small_gain_theorem} and is omitted  here due to lack of space.
	
	Finally, we provide an approach to compositionally construct an SSF for an interconnected POMDP $\Sigma$ based on SPSFs of its subsystems. Note that the constructed SSF is one of the main ingredients used in Theorem \ref{theorem_bound_xhat}. First, we raise the following small-gain assumption.
	\begin{assumption}\label{phis1}
		Assume that $\mathcal{K}_\infty$ functions $\mu_{ij}$ defined as 
		\begin{equation*}
			\bar{\mu}_{ij}(s):=
			\begin{cases}
				\bar{\mu}_i(s),\ \ \ \ \ \ \ \ \ \ \ &\text{\small if } \ i=j,
				\\\varrho_i(\varepsilon_j^{-1}(s)), \ \ \ \ \ \ \ \  \ \ \ \ \ \ \ \ \ & \text{\small if } \ i\neq j,
			\end{cases} 
		\end{equation*}
		satisfy
		\begin{equation}\label{eq:mu_ij_o}
			\bar{\mu}_{i_1 i_2}\circ \bar{\mu}_{i_2 i_3}\circ \cdots \circ\bar{\mu}_{i_{r-1} i_r}\circ \bar{\mu}_{i_r i_1}<\mathcal{I}_d,
		\end{equation}
		for all sequences $(i_1,\ldots,i_r)\in \{1,\ldots,N\}^r$ and $r\in\{1,\ldots,N\}$.
	\end{assumption}
	The small-gain condition \eqref{eq:mu_ij_o} implies the existence of $\mathcal{K}_\infty$ functions $\zeta_i>0$ \cite[Theorem 5.5]{ruffer2010monotone}, satisfying
	\begin{equation}\label{eq:zeta_mu_ij}
		\max_{i,j}\big{\{}\zeta_i^{-1}\circ\bar{\mu}_{ij}\circ\zeta_j\big{\}}<\mathcal{I}_d, \ \ \ \ \ \ \ i,j=\{1,\ldots,N\}. 
	\end{equation}
	In the next proposition, we show that if  Assumption~\ref{phis1} holds  and $\max_i\zeta_i^{-1}$ is concave, then we can compositionally construct an SSF for an interconnected system based on SPSFs of its subsystems.
	\begin{proposition}\label{phis2}
		Consider an interconnected POMDP $\Sigma=\mathcal{I}(\Sigma_i,\ldots,\Sigma_N)$ induced by $N\in\mathbb{N}_{\geq 1}$  subsystems $\Sigma_i$. Suppose that for each $\Sigma_i$ there exits an estimator $\widehat{\Sigma}_i$ together with a corresponding SPSF $\mbox{\boldmath$\phi$}_i(x_i,\hat{x}_i)$. If  Assumption~\ref{phis1} holds and $\max_i\zeta_i^{-1}$ for $\zeta_i$ as in \eqref{eq:zeta_mu_ij} is concave, then the function $\mbox{\boldmath$\phi$}(x,\hat{x})$ defined as 
		\[
		\mbox{\boldmath$\phi$}(x,\hat{x}) := \max_i\big\{\zeta_i^{-1}(\mbox{\boldmath$\phi$}_i(x_i,\hat{x}_i))\big\},
		\]
		is an SSF from $\widehat{\Sigma}=\mathcal{I}(\widehat{\Sigma}_i,\ldots,\widehat{\Sigma}_N)$ to $\Sigma=\mathcal{I}(\Sigma_i,\ldots,\Sigma_N)$, as defined in  Definition~\ref{def:phi_def}, with 
		\begin{align*}
			\mu(s)&= 	\max_{i,j}\big{\{}\zeta_i^{-1}\circ\bar{\mu}_{ij}\circ\zeta_j(s)\big{\}}, \ \ \ i,j=\{1,\ldots,N\},\\
			c &= \max_i \zeta_i^{-1}(\bar{c}_i).	
		\end{align*}
	\end{proposition}
	The proof of Proposition \ref{phis2} follows the same reasoning as that of  Theorem \ref{small_gain_theorem} and is omitted  here.
	
	\section{Computation of LCBF}\label{sos_section}
	In this subsection, we provide a systematic approach to search for LCBFs and the corresponding control policies for subsystems. The proposed approach is based on the sum-of-squares (SOS) optimization problem~\cite{parrilo2003semidefinite}, in which LCBF is restricted to be non-negative which can be written as a sum of squares of different polynomials. To do so, we need to raise the following assumption.
	
	\begin{assumption}\label{assumption:sos}
		The POMDP $\Sigma=(X,U,W,\varsigma_{1},f,Y_{1},Y_{2},h_{1},h_{2},\varsigma_{2})$ has a continuous state set $X\subseteq \mathbb{R}^{n}$ and continuous external and internal input sets $U\subseteq \mathbb{R}^{m}$ and $W\in\mathbb{R}^{p}$. Moreover, the transition map $f:X\times U\times W \times V_{\varsigma_{1}}\to X$ is a polynomial function of its arguments. We also assume that the internal
		output map $h_{1}:X\to Y_{1}$ and $\mathcal{K}_\infty$ functions $\alpha$ and $\rho$ are polynomial.
	\end{assumption}
	Under Assumption \ref{assumption:sos}, one can reformulate conditions of Definition \ref{definition_barrier} and Definition \ref{definition_barrier_xhat} to an SOS optimization problem in order to search for a polynomial LCBF $\mathcal{B}_i(\cdot,\cdot)$ and $\mathcal{B}_i(\cdot)$, and their corresponding control policies. In the following Lemmas, SOS formulations are provided.
	\begin{lemma}\label{lemma_sos}
		Suppose Assumption \ref{assumption:sos} holds and sets $X_a, X_b, X, W$  can be defined by vectors of polynomial inequalities $X_{a}=\{x\in \mathbb{R}^{n}\mid g_{a}(x)\geq 0\}, X_{b} =\{x\in \mathbb{R}^{n}\mid g_{b}(x)\geq 0\}, X=\{x\in \mathbb{R}^{n}\mid g(x)\geq 0\}$, and $W=\{w\in\mathbb{R}^{p}\mid g_{w}(w)\geq 0\}$, where the inequalities are defined element-wise. Suppose there exists a sum-of-square polynomial $\mathcal{B}(x,\hat{x})$, constants $\bar{\gamma},\tilde{\psi}\in\mathbb{R}_{\geq 0},\bar{\lambda}\in\mathbb{R}_{>0}$, $0<\tilde{\kappa}<1$, functions $\alpha\in \mathcal{K}_\infty,\tilde{\rho}\in\mathcal{K}_\infty\cup\{0\}$, polynomials $l_{\upsilon_{j}}(\hat{x},\hat{w})$ corresponding to the $j^{\text{th}}$ input in $\upsilon(k)=(\upsilon_{1}(k),\upsilon_{2}(k),\ldots,\upsilon_{m}(k))\in U\subseteq\mathbb{R}^{m}$, and vectors of sum-of-squares polynomials $l_{z}(x),\hat{l}_{z}(\hat{x})$ for $z\in \{0,1,2,3\}$, and $l_{w}(w),\hat{l}_{w}(\hat{w})$, of appropriate dimensions such that the following expressions are sum-of-square polynomials:
		\begin{equation}\label{eq:sos_cond_1}\begin{aligned}
				&\mathcal{B}(x,\hat{x})-\begin{bmatrix}l_{0}^T(x) \  \hat{l}_{0}^T(\hat{x})\end{bmatrix}\begin{bmatrix}
					g(x)\\g(\hat{x})\end{bmatrix}-\alpha(
				\begin{bmatrix} h_{1}(x)\\h_{1}(\hat{x})\end{bmatrix}^T\begin{bmatrix} h_{1}(x)\\h_{1}(\hat{x})\end{bmatrix}),
		\end{aligned}\end{equation}
		\begin{equation}\label{eq:sos_cond_2}
			-\mathcal{B}(x,\hat{x})-\begin{bmatrix}l_{1}^T(x) \ \ \hat{l}_{1}^T(\hat{x})\end{bmatrix}\begin{bmatrix}
				g_{a}(x)\\g_{a}(\hat{x})\end{bmatrix}+\bar{\lambda},
		\end{equation}
		\begin{equation}\label{eq:sos_cond_3}
			\mathcal{B}(x,\hat{x})-\begin{bmatrix}l_{2}^T(x) \ \ \hat{l}_{2}^T(\hat{x})\end{bmatrix}\begin{bmatrix}
				g_{b}(x)\\g(\hat{x})\end{bmatrix}+\bar{\gamma},
		\end{equation}
		\begin{equation}\label{eq:sos_cond_4}\begin{aligned}
				-\mathbb{E}&\Big{[}\mathcal{B}\big{(}
				f(x(k),\upsilon(k),w(k),\varsigma_{1}(k)), \hat{f}(\hat{x}(k),\upsilon(k),\hat{w}(k),y_{2}(k)) \big{)},
				\big|\, x(k), \hat{x}(k), \upsilon(k), w(k), \hat{w}(k)
				\Big{]} 
				+\tilde{\kappa}\mathcal{B}(x(k),\hat{x}(k))
				\\&
				+\tilde{\rho}(
				\frac{\begin{bmatrix}
						w(k)\\\hat{w}(k)\end{bmatrix}^T\begin{bmatrix}
						w(k)\\\hat{w}(k)\end{bmatrix}}{2p}
				)
				+\tilde{\psi}\sum_{j=1}^{m}(\upsilon_{j}(k)-l_{\upsilon_{j}}(\hat{x}(k),\hat{w}(k)))
				-\begin{bmatrix}
					l_{3}^T(x(k)) \ \ \hat{l}_{3}^T(\hat{x}(k)
				\end{bmatrix}
				\begin{bmatrix}
					g(x(k))\\g(\hat{x}(k))
				\end{bmatrix}
				\\&-
				\begin{bmatrix}
					l_{w}^T(w(k)) \ \ \hat{l}_{w}^T(\hat{w}(k))
				\end{bmatrix}
				\begin{bmatrix}
					g_{w}(w(k))\\g_{w}(\hat{w}(k))
				\end{bmatrix}\!,\\
			\end{aligned}
		\end{equation}
		where $p$ is the dimension of the internal inputs $w$ and $\hat{w}$. Then $\mathcal{B}(x,\hat{x})$ satisfies conditions \eqref{eq:barrier_w_1}-\eqref{eq:barrier_hat_w_4} in Definition \ref{definition_barrier} and $\upsilon(k)=[l_{\upsilon_{1}}(\hat{x}(k),\hat{w}(k));\ldots;l_{\upsilon_{m}}(\hat{x}(k)\hat{w}(k))]$ is the corresponding safety controller, with
		\begin{align*}
			\bar\kappa=&\mathcal{I}_d-(\mathcal{I}_d-\pi_{1})\circ(\mathcal{I}_d-\tilde{\kappa}),
			\\\rho=&(\mathcal{I}_d+\pi_{2})\circ(\mathcal{I}_d-\tilde{\kappa})^{-1}\circ\pi_{1}^{-1}\circ\pi_{3}\circ\tilde{\rho},
			\\\bar\psi=&(\mathcal{I}_d+\pi_{2}^{-1})\circ(\mathcal{I}_d-\tilde{\kappa})^{-1}\circ\pi_{1}^{-1}\circ\pi_{3}\circ(\pi_{3}-\mathcal{I}_d)^{-1}\circ(\tilde{\psi}),
		\end{align*}
		where $\pi_{1},\pi_{2},\pi_{3}$ being some arbitrarily chosen $\mathcal{K}_\infty$ functions so that $(\mathcal{I}_d-\pi_{1})\in\mathcal{K}_{\infty}$, and $(\pi_{3}-\mathcal{I}_d)\in\mathcal{K}_{\infty}$.
	\end{lemma}
	The proof follows the same argument as in \cite[Lemma 5.9]{jagtap2020compositional}, and is omitted here due to lack of space.
	
	\begin{remark}\label{remark_norm}
		Inequalities \eqref{eq:barrier_w_1} and \eqref{eq:barrier_w_4} consider infinity norms over $\begin{bmatrix} h_{1}(x);h_{1}(\hat{x}) \end{bmatrix}$ and $\begin{bmatrix}w;\hat{w}\end{bmatrix}$, respectively. Since such norms cannot be expressed as polynomials, we convert infinity norms to Euclidean ones and that is the reason constant $2p$ appears as a denominator in \eqref{eq:sos_cond_4}.
	\end{remark}
	
	We now state another lemma for the computation of LCBF as  in Definition \ref{definition_barrier_xhat}.

	\begin{lemma}\label{lemma_sos_xhat}
		Suppose Assumption \ref{assumption:sos} holds and sets $X_a, X_b^\epsilon, X, W, Y_2$  can be defined by vectors of polynomial inequalities $X_{a}=\{x\in \mathbb{R}^{n}\mid g_{a}(x)\geq 0\}, X_{b}^\epsilon =\{x\in \mathbb{R}^{n}\mid g_{b}^\epsilon(x)\geq 0\}, X=\{x\in \mathbb{R}^{n}\mid g(x)\geq 0\}$, $W=\{w\in\mathbb{R}^{p}\mid g_{w}(w)\geq 0\}$, and $Y_{2}=\{y_{2}\in\mathbb{R}^{q}\mid g_{y}(y_{2})\geq 0\}$
		where the inequalities are defined element-wise. Suppose there exists a sum-of-square polynomial $\mathcal{B}(x)$, constants $\bar{\gamma},\tilde{\psi}\in\mathbb{R}_{\geq 0},\bar{\lambda}\in\mathbb{R}_{>0}$, $0<\tilde{\kappa}<1$, functions $\alpha\in \mathcal{K}_\infty,\tilde{\rho}\in\mathcal{K}_\infty\cup\{0\}$, polynomials $l_{\upsilon_{j}}(\hat{x},\hat{w})$ corresponding to the $j^{\text{th}}$ input in $\upsilon(k)=(\upsilon_{1}(k),\upsilon_{2}(k),\ldots,\upsilon_{m}(k))\in U\subseteq\mathbb{R}^{m}$, and vectors of sum-of-squares polynomials $l_{z}(x)$ for $z\in \{0,1,2\}$, $\hat{l}_{3}(\hat{x})$, $\hat{l}_{w}(\hat{w})$ and $l_{y}(y_{2})$ of appropriate dimensions such that the following expressions are sum-of-square polynomials:

		\begin{align}\label{eq:sos_cond_1_xhat}
			&\mathcal{B}(x)-l_{0}^T(x) g(x)-\alpha( h_{1}(x)^Th_{1}(x)	
			),
			\\-&\mathcal{B}(x)- l_{1}^T(x)g_{a}(x)+\bar{\lambda},\label{eq:sos_cond_2_xhat}
			\\&\mathcal{B}(x)- l_{2}^T(x)g_{b}^\epsilon(x)+\bar{\gamma},\label{eq:sos_cond_3_xhat}
		\end{align}
		
		\begin{align}\notag
			-\mathbb{E}&\Big{[}\mathcal{B}\big{(}
			\hat{f}(\hat{x}(k),\upsilon(k),\hat{w}(k),y_{2}(k))
			\big{)}\mid \hat{x}(k), \upsilon(k), \hat{w}(k)\Big{]}
			+\tilde{\kappa}\mathcal{B}(\hat{x}(k))
			+\tilde{\rho}(\frac{
				\hat{w}^T(k)\hat{w}(k)}{p})
			+\tilde{\psi}
			\\&	-\sum_{j=1}^{m}(\upsilon_{j}(k)-l_{\upsilon_{j}}(\hat{x}(k),\hat{w}(k)))
			- \hat{l}_{3}^T(\hat{x}(k))
			g(\hat{x}(k))
			- \hat{l}_{w}^T(\hat{w}(k))g_{w}(\hat{w}(k))
			-l_{y}^T(y_{2}(k))g_{y}(y_{2}(k)),\label{eq:sos_cond_4_xhat}
		\end{align}
		where $p$ is the dimension of the internal input $w$.
		Then $\mathcal{B}(\hat{x})$ satisfies conditions \eqref{eq:barrier_hat_w_1}-\eqref{eq:barrier_hat_w_4} in Definition \ref{definition_barrier_xhat} and $\upsilon(k)=[l_{\upsilon_{1}}(\hat{x}(k),\hat{w}(k));\ldots;l_{\upsilon_{m}}(\hat{x}(k),\hat{w}(k))]$  is the corresponding safety controller, where $\bar\kappa,\rho,\bar\psi$ can be acquired based on $\tilde{\kappa},\tilde{\rho},\tilde{\psi}$ similar to Lemma~\ref{lemma_sos}.
	\end{lemma}
	\begin{remark}
		In order to compute the sum-of-square polynomials $\mathcal{B}(x,\hat{x})$ and $\mathcal{B}(x)$ fulfilling reformulated conditions \eqref{eq:sos_cond_1}-\eqref{eq:sos_cond_4}, and \eqref{eq:sos_cond_1_xhat}-\eqref{eq:sos_cond_4_xhat}, one can employ existing software tools such as SOSTOOLS \cite{prajna2002introducing} together with a semidefinite programming solver such as SeDuMi \cite{sturm1999using}.
	\end{remark}
	
	\section{Case Study}
\begin{figure}
		\centering
		\includegraphics[scale=0.15]{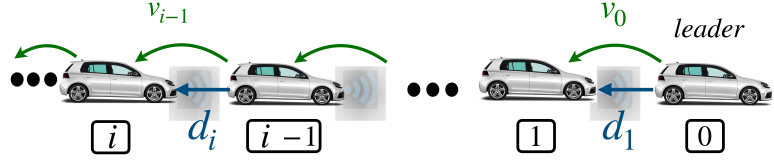} 	\vspace{-2cm}
		\caption{Platoon model for $N=1000$ vehicles.}
		\label{fig:cars}
		\vspace{-0.5cm}
	\end{figure}

\begin{figure}
		\centering
		\includegraphics[scale=0.26]{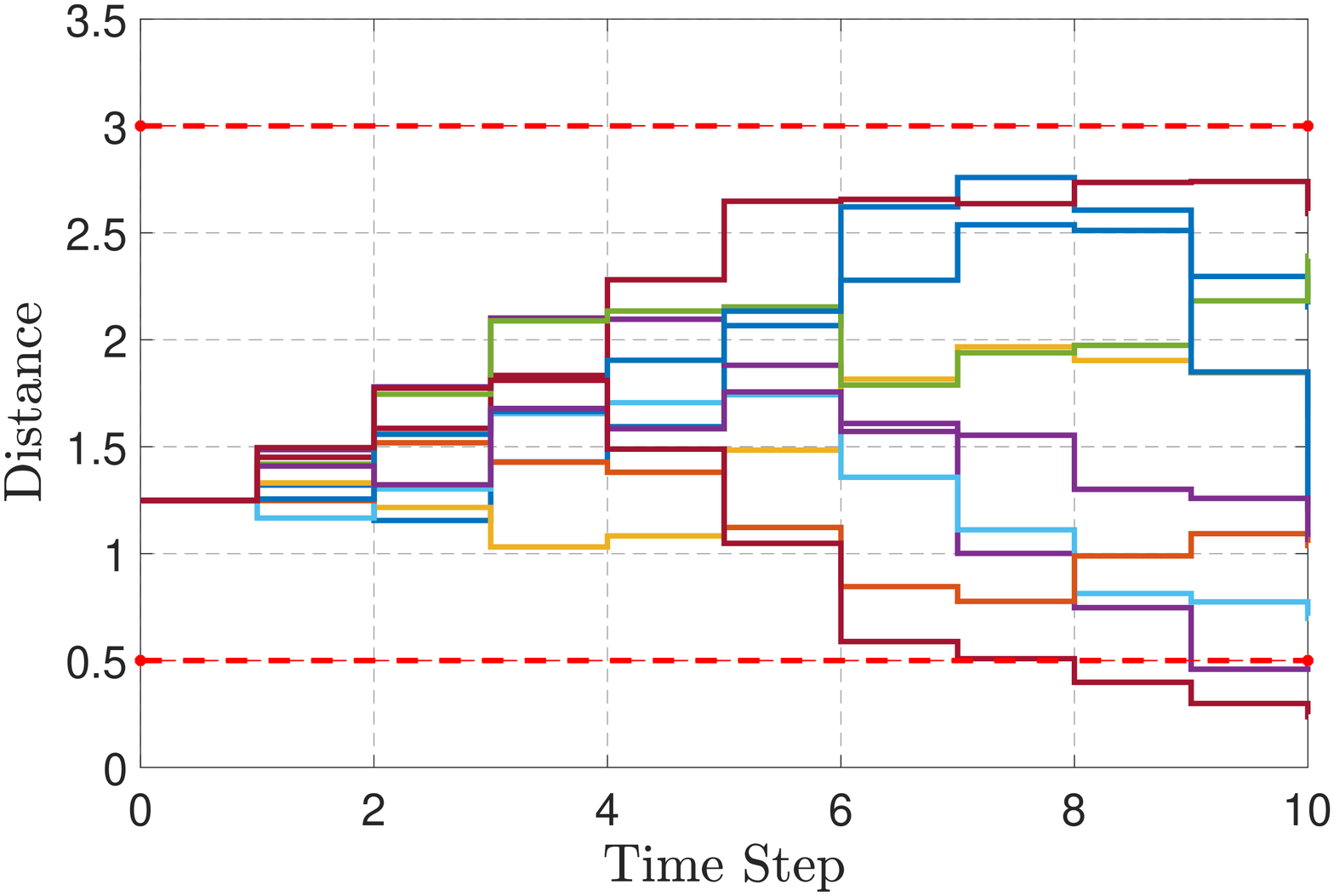}
		\includegraphics[scale=0.26]{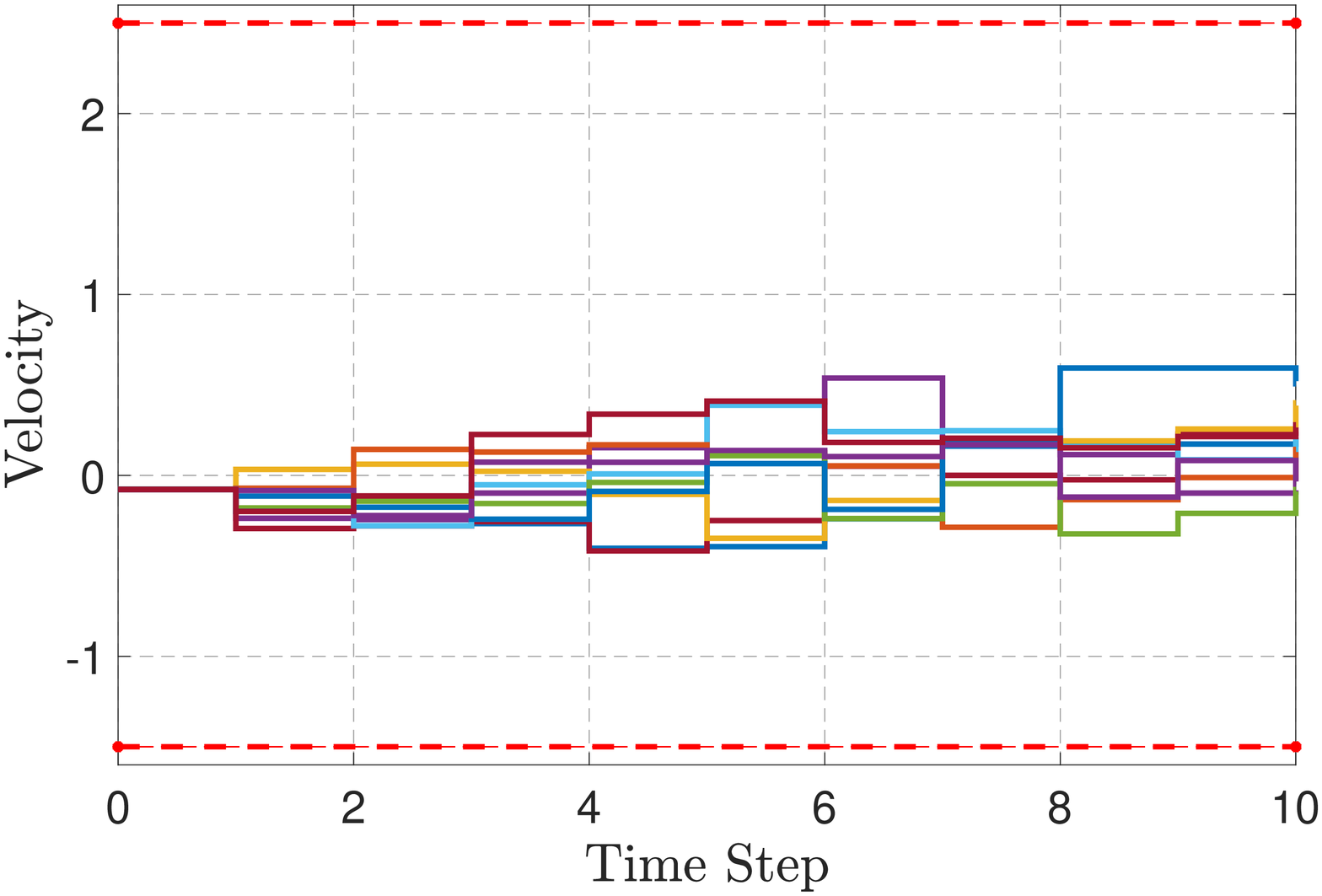}
		\includegraphics[scale=0.26]{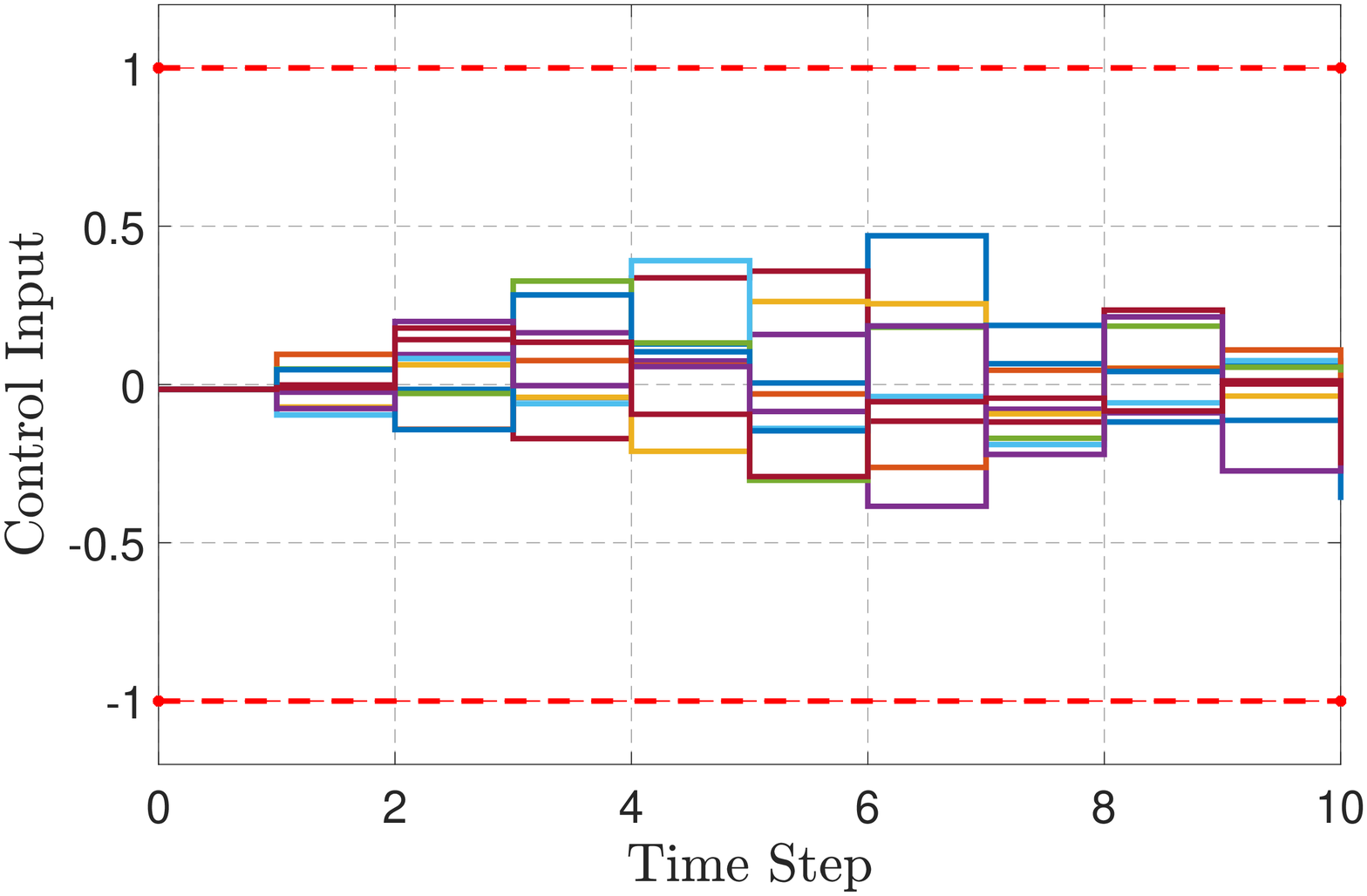}
		\caption{Closed-loop state (distance and velocity) and input trajectories of a representative vehicle with different noise realizations in a network of 1000 vehicles under controller \eqref{eq:u1_b1}.}
		\label{fig:1}
	\end{figure}

	In this section, we illustrate our proposed results by applying them to an adaptive cruise control $($ACC$)$ system consisting of $N$ vehicles in a platoon $($see ~Fig. \ref{fig:cars}$)$. This model is adapted from \cite{sadraddini2017provably}. The evolution of states can be described by the interconnected PO-dt-SCS
	\begin{equation*}
		\Sigma:\begin{cases}
			x(k+1)=\bar{A}x(k)+\bar{B}\upsilon(k)+\varsigma_{1}(k),
			\\y(k)=\bar{C}x(k)+\varsigma_{2}(k),
		\end{cases}
	\end{equation*}
	where $\bar{A}$ is a block matrix with diagonal blocks $A$, and off-diagonal blocks $A_{i(i-1)}=A_w, i\in\{2,\ldots,N\}$, where
	\[A=\begin{bmatrix}
		1 & -1\\ 0  &1
	\end{bmatrix}\!, \quad A_w=\begin{bmatrix}
		0 & \tau\\ 0& 0
	\end{bmatrix}\!,
	\]
	with $\tau=0.01$ being the interconnection degree, and all other off-diagonal blocks being zero matrices of appropriate dimensions. Moreover, $\bar{B}$ is a partitioned matrix with main diagonal blocks $B=[0~\!;1]$, and all other off-diagonal blocks being zero matrices of appropriate dimensions. The matrix $\bar{C}$ is a partitioned matrix with main diagonal blocks $C=[1~\!;0]^T$ and all other off-diagonal blocks being zero matrices of appropriate dimensions. Moreover, $x(k)=[x_1(k);\ldots;x_N(k)]$, $\upsilon(k)=[\upsilon_1(k);\ldots;\upsilon_N(k)]$,
	$\varsigma_{1}(k)=[\varsigma_{1_1}(k);\ldots;\varsigma_{1_N}(k)]$, and $\varsigma_{2}(k)=[\varsigma_{2_1}(k);\ldots;\varsigma_{2_N}(k)]$. Let us consider each individual vehicle $\Sigma_i$ described as 
	\begin{equation*}  
		\Sigma_i:	\begin{cases}
			x_i(k+1)=Ax_i(k)+B\upsilon_i(k)+A_ww_i(k)+\varsigma_{1_i}(k),
			\\y_{1_i}(k)=C_1x_i(k),
			\\y_{2_i}(k)=C_2x_i(k)+\varsigma_{2_i},
		\end{cases}
	\end{equation*}
	where $y_{1_i}(k)=y_{1_{i(i+1)}}(k)=C_1x_i(k),i\in\{1,\ldots N\}$, (with $C_1=[0~\!;1]$ and $y_{1_{N(N+1)}}=0$) and $C_2=C$.
	One can readily verify that $\Sigma=\mathcal{I}(\Sigma_i,\ldots,\Sigma_N)$, where $w_i(k)=[0;w_{i(i-1)}(k)],i\in\{1,\ldots,N\}$, (with $w_{i(i-1)}=y_{1_{(i-1)i}}=C_1 x_{i-1},w_{1,0}=0$). The state of the $i$-th vehicle is defined as $x_i=[d_i;\mathrm{v}_i]$, for $i\in\{1,\ldots,N\}$, where $d_i$ denotes the relative distance between the vehicle $i$ and its proceeding vehicle $i-1$ (the $0$-$th$ vehicle represents the leader), $\mathrm{v}_i$ is its velocity in the leader’s frame, and $\upsilon_i\in [
	-1 , 1]$ is the bounded control input. The overall control objective in ACC is for each vehicle to adjust its speed in order to maintain a safe distance from the vehicle ahead \cite{jahanshahi2018attack}.
	For the system $\Sigma_i$, we design a proper estimator of the following form 
	\begin{equation*}
		\hat{\Sigma}_i:	\begin{cases}\begin{aligned}
				\hat{x}_i(k+1)=A\hat{x}_i(k)+B\upsilon_i(k)+A_w\hat{w}_i(k)+K(y_{2_i}(k)-C_2\hat{x}_i(k)),	\end{aligned}
			\\\hat{y}_{1_i}(k)=C_1\hat{x}_i(k),
		\end{cases}
	\end{equation*}
	where $K=[ 1.7; -0.72]$ is the estimator gain. We consider a network of $N=1000$ vehicles where the regions of interest for each vehicle are $X\in[0,3.5]\times[-2,3 ] $, $X_a\in[1,1.5]\times[-0.4,0.4 ]$, and $X_b\in[0,0.5]\times[-2,-1.5 ] \cup [3,3.5]\times[2.5 ,3 ] $. Now, for each vehicle we compute LCBFs while compositionally synthesizing safety controllers for a bounded-time horizon. We construct LCBFs using the two methods introduced in Section~\ref{CBD_section} and employ the software SOSTOOLS to search for LCBFs as described in Section~\ref{sos_section}.
	According to Section~\ref{Barrier_1}, we compute the LCBF $\mathcal{B}_i(x_i,\hat{x}_i)$ of an order $4$ and its corresponding controller as the following:
	\begin{equation}\label{eq:u1_b1}		
		\upsilon_i= 0.06\hat{d}_i -0.7\hat{\mathrm{v}}_i + 0.02\hat{\mathrm{v}}_{i-1} -0.07,
	\end{equation}
	for $i\in\{1,\ldots,N\}$. 
	Moreover, the corresponding constants and functions in Definition~\ref{definition_barrier} are quantified as $\alpha_i(s)=10^{-5}s, s\in \mathbb{R}_{\geq 0}, \bar{\gamma}_i=0.12,
	\bar{\lambda}_i=1,\bar{\kappa}_i=0.95, \rho_i(s)=2\times 10^{-8}s, s\in \mathbb{R}_{\geq 0}, \bar{\psi}_i=0.001$. 
	Now, we check the small gain condition \eqref{eq:k_ij_o} that is
	required for the compositionality result. By taking $\sigma_i(s)=s$, $i\in\{1,\ldots,N\}$, the condition \eqref{eq:k_ij_o}, and as a result the condition \eqref{eq:sigma_k_ij} are always satisfied without any restriction on the number of vehicles. Hence, $\mathcal{B}(x,\hat{x})=\max_i\mathcal{B}_i(x_i,\hat{x_i})$ is a CBF for $\Sigma$ satisfying conditions in Definition~\ref{definition_barrier_no_w} with $\gamma=0.12, \lambda=1, \kappa=0.95, \psi=0.001$. By employing Theorem~\ref{theorem_bound}, one can guarantee that states of the interconnected system staring from $X_a$ remain in the safe set $X\backslash X_b$ within the time horizon $T_d=10$ with a probability of at least $87.12\%$. Closed-loop state and input trajectories of a representative vehicle with different noise realizations are illustrated in Fig. \ref{fig:1} with only 10 trajectories.
	
    	\begin{figure}
    	\centering
    	\includegraphics[scale=0.26]{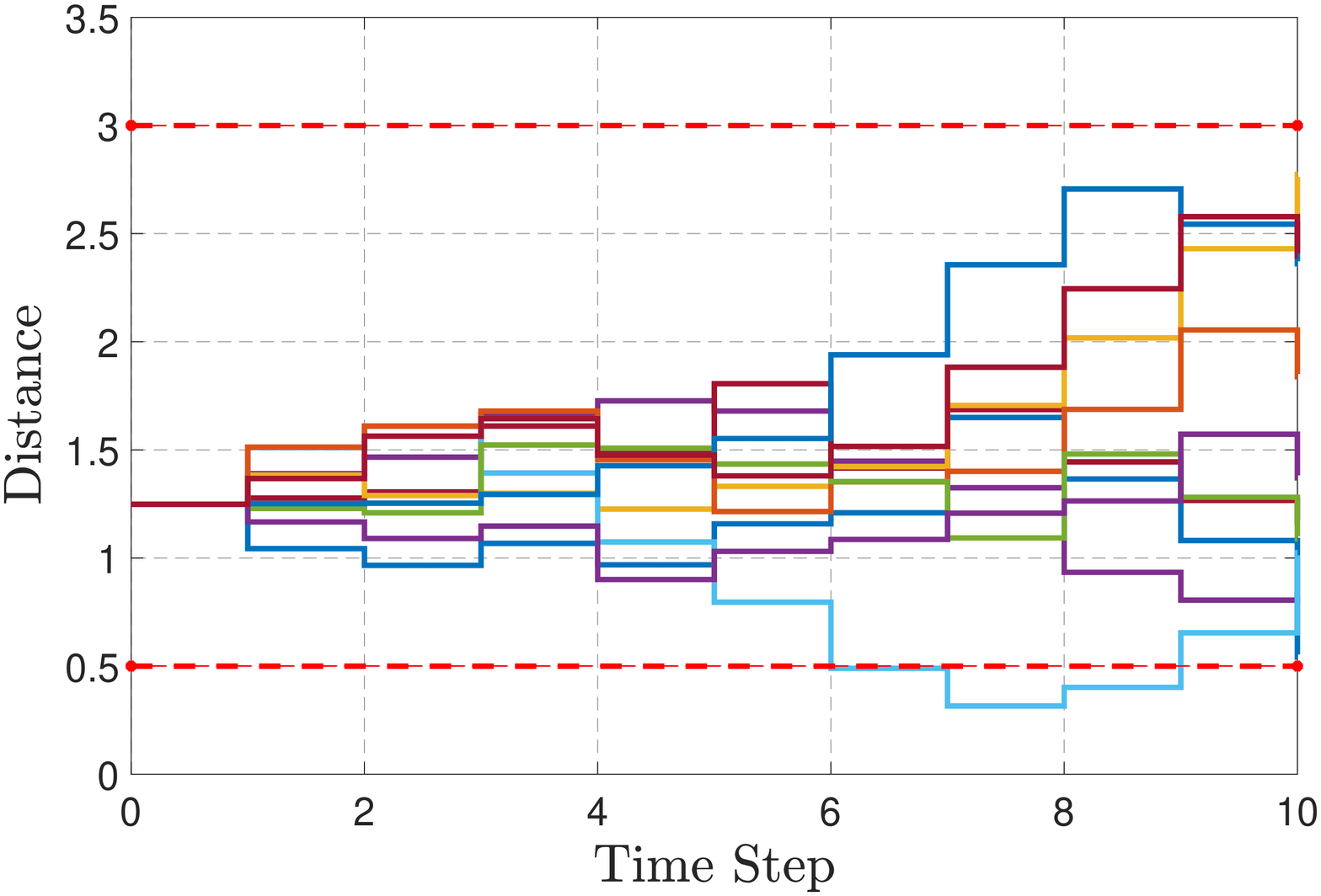}
    	\includegraphics[scale=0.26]{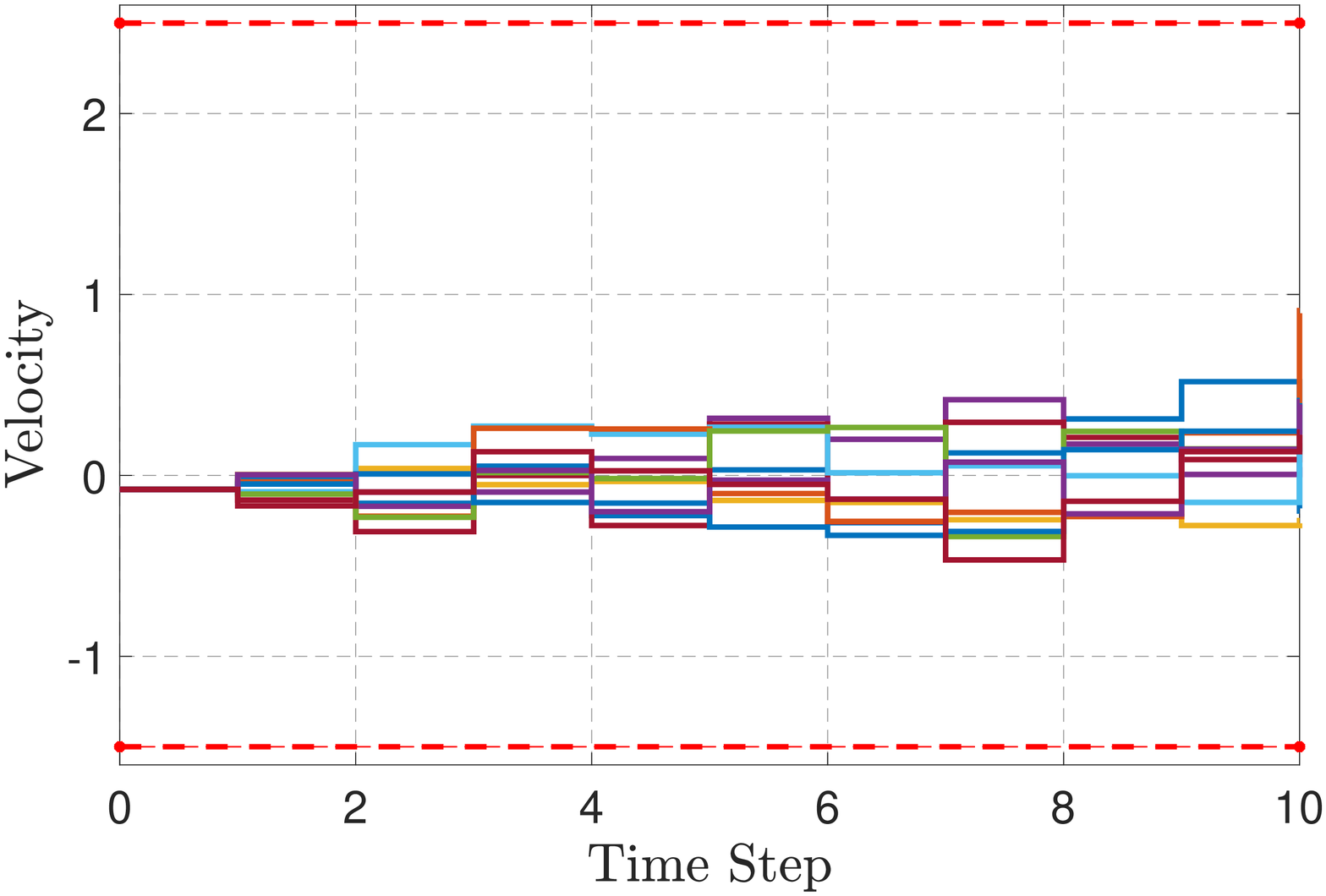}
    	\includegraphics[scale=0.26]{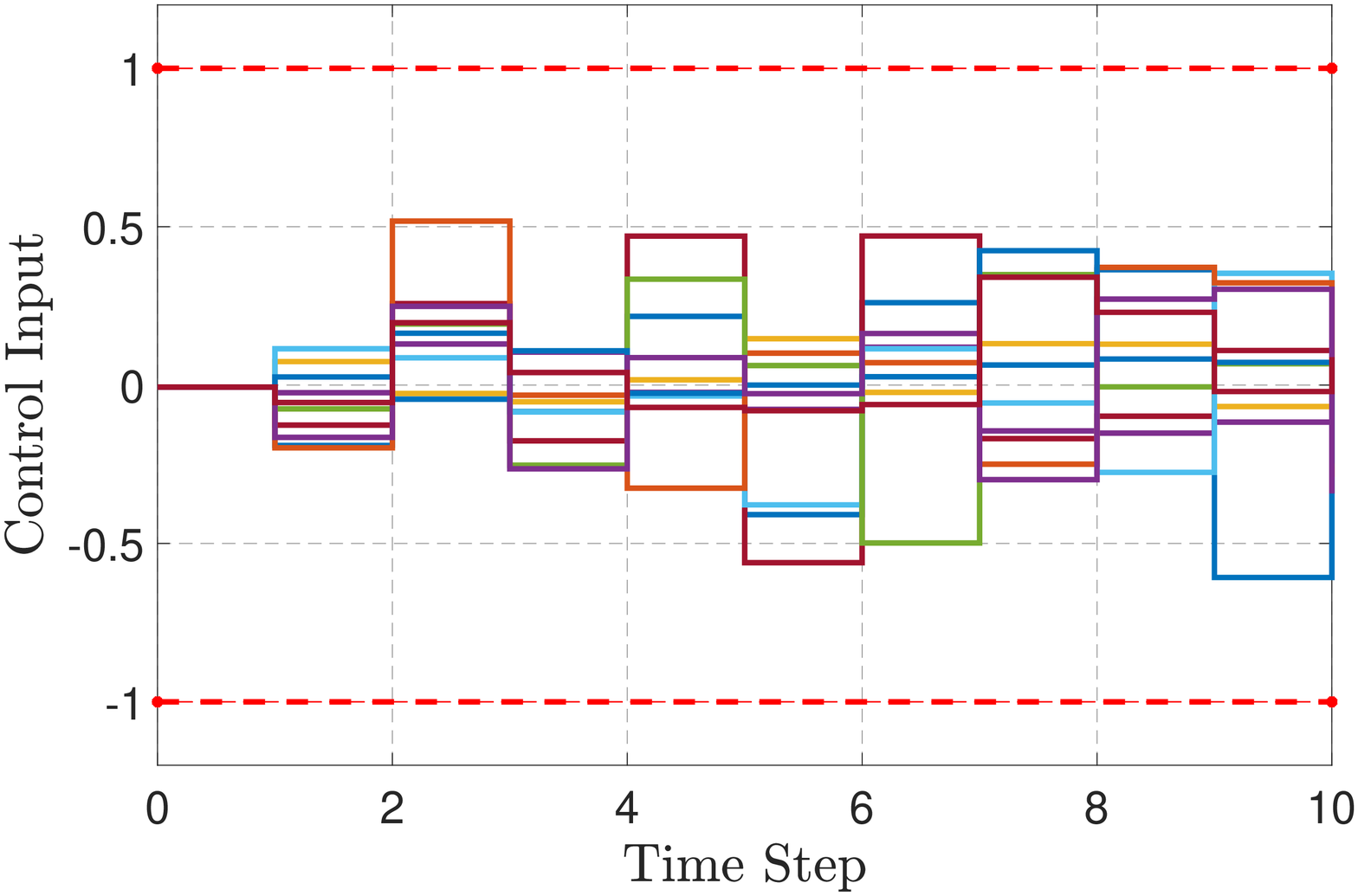}
    	\caption{Closed-loop state (distance and velocity) and input trajectories of a representative vehicle with different noise realizations in a network of 1000 vehicles under controller \eqref{eq:u1_b2}.}
    	\label{fig:2}
    \end{figure}

	We now construct the LCBF $\mathcal{B}_i(\hat{x_i})$ of an order $4$ for the estimator, as described in Section~\ref{Barrier_2}, and compute its corresponding controller as
	\begin{equation}\label{eq:u1_b2}
		\upsilon_i= 0.09\hat{d}_i -\hat{\mathrm{v}}_i + 0.03\hat{\mathrm{v}}_{i-1} -0.09,
	\end{equation}
	for $i\in\{1,\ldots,N\}$. The corresponding constants and functions in Definition~\ref{definition_barrier_xhat} are quantified as $\alpha_i(s)=10^{-5}s, s\in \mathbb{R}_{\geq 0}, \bar{\gamma}_i=0.12,
	\bar{\lambda}_i=1,\bar{\kappa}_i=0.95, \rho_i(s)=2\times 10^{-8}s, s\in \mathbb{R}_{\geq 0}, \bar{\psi}_i=0.001$. Similar to the first method, we check the small gain condition \eqref{eq:k_ij_o} for the compositionality result. By taking $\sigma_i(s)=s$, $i\in\{1,\ldots,N\}$, the condition \eqref{eq:k_ij_o}, and as a result the condition \eqref{eq:sigma_k_ij} are both satisfied. Hence, $\mathcal{B}(\hat{x})=\max_i\mathcal{B}_i(\hat{x_i})$ is a CBF for $\Sigma$ satisfying conditions in Definition~\ref{definition_barrier_no_w_xhat} with $\gamma=0.12, \lambda=1, \kappa=0.95, \psi=0.001$. By employing the result of Theorem~\ref{theorem_bound}, one can guarantee that the states of the estimator staring from $X_a$ will not reach $X_b^\epsilon$ within the time horizon $T_d=10$ with a probability of at least $ 87.12\%$. Now, in order to compute the exit probability bound for the interconnected system, we search for an SPSF of a quadratic form $\mbox{\boldmath$\phi$}_i(x_i,\hat{x}_i)=(x_i-\hat{x}_i)^TM(x_i-\hat{x}_i)$, where $M$ is a positive-definite matrix. Since the dynamic of the system is linear, the conditions in Definition~\ref{def:phi_def_w} reduce to solving the following matrix inequality:
	\[(1+2/\tilde{\pi})(A-KC_2)^TM(A-KC_2)\leq \bar{\mu}M, \]
	where $K$ is the estimator gain, and $\tilde{\pi}>0$. By using the tool YALMIP \cite{lofberg2004yalmip}, we compute $M$ as 
	\[
	M=
	\begin{bmatrix}
		0.0257   & 0.0259\\
		0.0259   &  0.0262
	\end{bmatrix},
	\]
	with $\tilde{\pi}=1$. The functions and constants associated with this SPSF are computed  by following the compositional construction method for linear systems introduced in \cite[Theorem 6.10]{lavaei2020compositional} as $\varepsilon(s)=0.3s^2,s\in \mathbb{R}_{\geq 0},  \bar{\mu}=0.4 ,\varrho(s)=0.002s^2, s\in \mathbb{R}_{\geq 0},  \bar{c} = 10^{-5}$. Hence, $\mbox{\boldmath$\phi$}(x,\hat{x})=\max_i\mbox{\boldmath$\phi$}_i(x_i,\hat{x}_i)$ is an SSF from $\widehat{\Sigma}$ to $\Sigma$ satisfying the conditions in Definition~\ref{def:phi_def} with $\varepsilon(s)=0.3s^2, s\in \mathbb{R}_{\geq 0}, \mu=0.4, c = 10^{-5}$, $\epsilon=0.01$. An upper bound of $3.61\%$ on the probability of the estimation accuracy is computed according to Theorem~\ref{thm:SSF} within the time horizon $T_d=10$. Employing Theorem~\ref{theorem_bound_xhat}, the probability that the solution process of the system starting from the initial region $X_a$ and not reaching $X_b$ is at least $ 83.51\%$. Closed-loop state and input trajectories of a representative vehicle with different noise realizations are illustrated in Fig.~\ref{fig:2}.
	
	\section{Conclusions}
	In this paper, we proposed a compositional approach based on control barrier functions for the synthesis of safety controllers for networks of POMDP by utilizing small-gain type reasoning. The proposed scheme provides an upper bound on the probability that the interconnected system reaches an unsafe region in a finite-time horizon. In this respect, we first quantified probability bounds without any prior information of the estimation accuracy. This is achieved by constructing local barrier functions over an augmented system composed of subsystems and their corresponding estimators. Alternatively, we formulated local barrier functions based on only estimators' dynamics and computed the exit probability by utilizing the probability bound on the estimation accuracy computed via notions of stochastic simulation functions. We finally demonstrated the effectiveness of our proposed results by applying them to an adaptive cruise control problem.
	
	\bibliographystyle{alpha}
	\bibliography{biblio}
	
	\section{Appendix}
	
	\begin{proof}\textbf{(Theorem \ref{theorem_bound})}
			According to condition \eqref{eq:barrier_no_w_2}, $X_b\times X\subseteq\{(x,\hat{x})\in X\times X \, | \ \mathcal{B}(x,\hat{x})\geq\lambda\}$. Then we have 
			\begin{equation}\label{eq:pr_inq}
				\begin{aligned}
					\mathbb{P}&\big{[}x_{a\upsilon}(k)\in X_b\wedge\hat{x}_{\hat{a}\upsilon}(k)\in X \ \text{for some} \  k\in[0,T_d] \ \big | \  a, \hat{a}, \upsilon \big{]}
					\\	& \leq\mathbb{P}\big{[}\sup_{0\leq k\leq T_d} 
					\mathcal{B}(x_{a\upsilon}(k),\hat{x}_{\hat{a}\upsilon}(k)) \geq \lambda
					\ \big | \ a,\hat{a}, \upsilon\big{]}\leq \delta.
				\end{aligned}
			\end{equation}
			The proposed bounds in \eqref{eq:delta_bound} follow directly by applying 
			\cite[Theorem 3, Chapter III]{kushner1965stability}
			to the above inequality and employing conditions \eqref{eq:barrier_no_w_3} and \eqref{eq:barrier_no_w_1}, respectively.
			Inequality \eqref{eq:pr_inq} is obtained by utilizing the result of \cite[Theorem 1]{kushner1967stochastic}.
			Now we get 
			\begin{align*}
				\mathbb{P}&\big{[}x_{a\upsilon}(k)\in X_b\wedge\hat{x}_{\hat{a}\upsilon}(k)\in X \ \text{for some} \ k\in[0,T_d] \ \big | \ a, \hat{a}, \upsilon \big{]}
				\\ &\leq\mathbb{P}\big{[}x_{a\upsilon}(k)\in X_b \ \text{for some}\  k\in[0,T_d] \ | \ a, \upsilon\big{]}
				\\ &~~~+
				\mathbb{P}\big{[}\hat{x}_{\hat{a}\upsilon}(k)\in X \ \text{for some} \ k\in[0,T_d] \ \big | \ \hat a, \upsilon\big{]}
				\\&~~~-
				\mathbb{P}\big{[}x_{a\upsilon}(k)\in X_b \vee \hat{x}_{\hat{a}\upsilon}(k)\in  X  \ \text{for some} \  k\in[0,T_d] \ \big | \ a, \hat{a}, \upsilon\big{]}\!.
			\end{align*}
			Since, the second and last terms trivially hold with probability 1, one has
			\begin{align*}
				\mathbb{P}&\big{[}x_{a\upsilon}(k)\in X_b\wedge\hat{x}_{\hat{a}\upsilon}(k)\in X \ \text{for some} \ k\in[0,T_d] \ \big | \ a, \hat{a},\upsilon\big{]}
				\\ & \leq\mathbb{P}\big{[}x_{a\upsilon}(k)\in X_b  \ \text{for some} \  k\in[0,T_d] \ \big | \  a, \upsilon \big{]}.
			\end{align*}
			Now, since the right term of the conjunction (i.e., $\wedge$) holds for all time, the inequality above becomes an equality and one gets
			$\mathbb{P}\big{[}x_{a\upsilon}(k)\in X_b \ \text{for some} \ k\in[0,T_d] \ \big | \  a,\upsilon\big{]} \leq \delta$
			which concludes the proof.
	\end{proof}
	
	\begin{figure*}[h!]
		\rule{\textwidth}{0.1pt}
		\begin{align}\notag
			\mathbb{E}&\Big{[}\mathcal{B}(	f(x(k),\upsilon(k),\varsigma_{1}(k)), \hat{f}(\hat{x}(k),\upsilon(k), y(k)) )\,\big |\, x(k), \hat{x}(k), \upsilon(k)\Big{]}
			\\\notag &=\mathbb{E}\Big{[}\max_i\Big{\{}\sigma_i^{-1}(\mathcal{B}_i(f_i(x_i(k),\!\upsilon_i(k),\!w_i(k),\!\varsigma_{1_i}(k)),\! \hat{f}_i(\hat{x}_i(k),\!\upsilon_i(k),\!\hat{w}_i(k),\!y_{2_i}(k))))
			\Big{\}}\,\big |\, x(k),\!\hat{x}(k),\!\upsilon(k),\! w(k),\!\hat{w}(k)\Big{]}
			\\\notag
			&\leq \max_i\Big{\{}\sigma_i^{-1}(\mathbb{E}\Big{[}\mathcal{B}_i(f_i(x_i(k),\!\upsilon_i(k),\!w_i(k),\varsigma_{1_i}(k)),\! \hat{f}_i(\hat{x}_i(k),\!\upsilon_i(k),\!\hat{w}_i(k),\!y_{2_i}(k)))\,\big |\, x(k),\!\hat{x}(k),\!\upsilon(k),\!w(k),\hat{w}(k)\Big{]})\Big{\}}
			\\\notag
			&	=\max_i\Big{\{}\sigma_i^{-1}(\mathbb{E}\Big{[}\mathcal{B}_i(f_i(x_i(k),\!\upsilon_i(k),\!w_i(k),\!\varsigma_{1_i}(k)),\! \hat{f}_i(\hat{x}_i(k),\!\upsilon_i(k),\!\hat{w}_i(k),\!y_{2_i}(k)))\big |x_i(k),\! \hat{x}_i(k),\!\upsilon_i(k),\!w_i(k),\!\hat{w}_i(k)\Big{]})\Big{\}}
			\\\notag
			&\leq\max_i\Big{\{}\sigma_i^{-1}(\max\{\bar{\kappa}_i(\mathcal{B}_i(x_i(k),\hat{x}_i(k))),\rho_i(\|
			\begin{bmatrix}
				w_i(k)\\\hat{w}_i(k)
			\end{bmatrix}
			\|^{^2}),\bar{\psi}_i\})\Big{\}}
			\\\notag &	=\max_i\Big{\{}\sigma_i^{-1}(\max\{\bar{\kappa}_i(\mathcal{B}_i(x_i(k),\hat{x}_i(k))),\rho_i(\max_{j,j\neq i}\| \begin{bmatrix}
				w_{ij}(k)\\\hat{w}_{ij}(k)
			\end{bmatrix}\|^{^2}),\bar{\psi}_i\})\Big{\}}
			\\\notag
			&=\max_i\Big{\{}\sigma_i^{-1}(\max\{\bar{\kappa}_i(\mathcal{B}_i(x_i(k),\hat{x}_i(k))),\rho_i(\max_{j,j\neq i}\| 
			\begin{bmatrix}
				y_{1_{ji}}(k)\\\hat{y}_{1_{ji}}(k)
			\end{bmatrix}\|^{^2}),\bar{\psi}_i\})\Big{\}}
			\\\notag
			&\leq\max_i\Big{\{}\sigma_i^{-1}(\max\{\bar{\kappa}_i(\mathcal{B}_i(x_i(k),\hat{x}_i(k))),\rho_i(\max_{j,j\neq i}\| 
			\begin{bmatrix}
				h_{1_j}(x_j(k))\\h_{1_j}(\hat{x}_j(k))
			\end{bmatrix}\|^{^2}),\bar{\psi}_i\})\Big{\}}
			\\\notag
			&\leq \max_i\Big{\{}\sigma_i^{-1}(\max\{\bar{\kappa}_i(\mathcal{B}_i(x_i(k),\hat{x}_i(k))),\rho_i(\max_{j,j\neq i}\{\alpha^{-1}_j(\mathcal{B}_j(x_j(k),\hat{x}_j(k)))\}),\bar{\psi}_i\})\Big{\}}
			\\\notag &	=\max_{i,j}\Big{\{}\sigma_i^{-1}(\max\{\bar{\kappa}_{ij}(\mathcal{B}_i(x_i(k),\hat{x}_i(k))),\bar{\psi}_i\})\Big{\}}
			=\max_{i,j}\Big{\{}\sigma_i^{-1}(\max\{\bar{\kappa}_{ij}\circ\sigma_j\circ \sigma_j^{-1}(\mathcal{B}_j(x_j(k),\hat{x}_j(k))),\bar{\psi}_i\})\Big{\}}
			\\\notag &
			\leq\max_{i,j,l}\Big{\{}\sigma_i^{-1}(\max\{\bar{\kappa}_{ij}\circ\sigma_j\circ \sigma_l^{-1}(\mathcal{B}_l(x_l(k),\hat{x}_l(k))),\bar{\psi}_i\})\Big{\}}
			\\\label{eq:inq_chain} 
			&=\max_{i,j}\Big{\{}\sigma_i^{-1}(\max\{\bar{\kappa}_{ij}\circ\sigma_j(\mathcal{B}(x(k),\hat{x}(k))),\bar{\psi}_i\})\Big{\}}
			=\max\{\kappa(\mathcal{B}(x(k),\hat{x}(k))),\bar{\psi}\Big{\}}.
		\end{align}
		\rule{\textwidth}{0.1pt}
	\end{figure*}

	\begin{proof}\textbf{(Theorem \ref{thm:SSF})}
			Since $\mbox{\boldmath{$\phi$}}$ is a stochastic pseudo-simulation function from $\widehat{\Sigma}$ to $\Sigma$, one has
			\begin{equation*}\begin{aligned}
					\mathbb{P}&\Big{[}\sup_{0\leq k\leq T_d}\|x_{a\upsilon}(k)-\hat{x}_{\hat{a} \upsilon}(k)\|\geq \epsilon \,\big|\, a,\hat{a}, \upsilon\Big{]}
					\\&=\mathbb{P}\Big{[}\sup_{0\leq k\leq T_d}\varepsilon(\|x_{a\upsilon}(k)-\hat{x}_{\hat{a} \upsilon}(k)\|)\geq \varepsilon(\epsilon)\,\big|\, a,\hat{a}, \upsilon\Big{]}
					\\&\leq \mathbb{P}\Big{[}\sup_{0\leq k\leq T_d}\mbox{\boldmath{$\phi$}}(x_{a \upsilon}(k),\hat{x}_{\hat{a}\upsilon}(k))\geq \varepsilon(\epsilon)\,\big|\, a,\hat{a}, \upsilon\Big{]}\leq \theta.
			\end{aligned}\end{equation*}
			The equality holds due to the fact that $\varepsilon$ is a $\mathcal{K}_\infty$ function. The second inequality holds based on the first condition of Definition \ref{def:phi_def}, and the last inequality follows from the result in \cite[Theorem 1]{kushner1965stability}.
	\end{proof}

	\begin{proof}\textbf{(Theorem \ref{small_gain_theorem})}
			We first show that conditions \eqref{eq:barrier_no_w_1} and \eqref{eq:barrier_no_w_2} in Definition~\ref{definition_barrier_no_w} hold. For any $(x,\hat{x})\in X_a\times X_a$, with $X_a=\prod_{i=1}^{N}X_{a_i}$, and from \eqref{eq:barrier_w_2}, we have
			\begin{align*}
				\mathcal{B}(x,\hat{x})=\max_i\big{\{}\sigma^{-1}_i(\mathcal{B}_i(x_i,\hat{x}_i))\big{\}}\leq \max_i
				\big{\{}\sigma^{-1}_i(\bar{\gamma}_i)\big{\}}=\gamma,
			\end{align*}
			and simply for any $(x,\hat{x})\in X_b\times X$, with $X_b=\prod_{i=1}^{N}X_{b_i}$, $X=\prod_{i=1}^{N}X_{i}$ and from \eqref{eq:barrier_w_3}, we have 
			\begin{align*}
				\mathcal{B}(x,\hat{x})=\max_i\big{\{}\sigma^{-1}_i(\mathcal{B}_i(x_i,\hat{x}_i))\big{\}}\geq \max_i
				\big{\{}\sigma^{-1}_i(\bar{\lambda}_i)\big{\}}=\lambda,
			\end{align*}
			satisfying conditions \eqref{eq:barrier_w_2} and \eqref{eq:barrier_w_3} with $\gamma=\max_i
			\big{\{}\sigma^{-1}_i(\bar{\gamma}_i)\big{\}}$ and  $\lambda=\max_i
			\big{\{}\sigma^{-1}_i(\bar{\lambda}_i)\big{\}}$. Moreover, $\lambda> \gamma$ according to \eqref{eq:gamma_einq_lambda}.
			Now we show that condition \eqref{eq:barrier_no_w_3} holds, as well. Let $\kappa(s)=\max_{i,j}\{\sigma_{j}^{-1}\circ\bar{\kappa}_{ij}\circ\sigma_j(s)\}$. It follows from \eqref{eq:sigma_k_ij} that $\kappa < \mathcal{I}_d$.  Since $\text{max}_i \sigma_i^{-1}$ is concave, one can readily acquire the chain of inequalities in \eqref{eq:inq_chain} using Jensen’s inequality. Hence, $\mathcal{B}$ is a CBF for the augmented system $\widetilde{\Sigma}=[\Sigma;\widehat{\Sigma}]$, which completes the proof.
	\end{proof}

\end{document}